\documentclass[preprint,10pt]{elsarticle}
\usepackage{amsfonts,amssymb,amsmath,empheq}
\usepackage{amsthm}
\usepackage{booktabs}
\usepackage{graphics}
\usepackage{graphicx,color}
\usepackage{latexsym}
\usepackage{subfig}
\usepackage{multirow}
\usepackage{natbib}
\usepackage{float}
\usepackage{cases}
\usepackage{bm}
\usepackage{color}
\usepackage{indentfirst}
\usepackage{epstopdf}
\usepackage{CJK}
\usepackage[colorlinks,
            linkcolor=red,
            anchorcolor=blue,
            citecolor=green
            ]{hyperref}
\numberwithin{equation}{section}

\newcommand{\blue}{\color{blue}}

\newcommand{\beqa}{\begin{eqnarray}}
\newcommand{\eeqa}{\end{eqnarray}}
\newcommand{\pr}{\partial}
\newcommand{\bx}{\mbox{\boldmath $x$}}
\newcommand{\bq}{\mbox{\boldmath $q$}}
\newcommand{\beq}{\begin{equation}}
\newcommand{\eeq}{\end{equation}}

\journal{}

\begin{document}

\begin{frontmatter}

\title{A discontinuous Galerkin method for wave propagation in orthotropic poroelastic media with memory terms}
\author[author1]{Jiangming Xie}
\ead{xiejm@cqu.edu.cn}
\author[author2]{M.Yvonne Ou}
\ead{mou@udel.edu}
\author[author3]{Liwei Xu}
\ead{xul@uestc.edu.cn}

\address[author1]{College of Mathematics and Statistics, Chongqing University, Chongqing, 401331, China}
\address[author2]{Department of Mathematical Sciences, University of Delaware, Newark, DE 19716, U.S.A}
\address[author3]{School of Mathematical Sciences, University of Electronic Science and Technology of China, Sichuan 611731, China}
\begin{abstract}
In this paper, we investigate wave propagation in orthotropic poroelastic media by studying the time-domain poroelastic equations. Both the low frequency Biot's (LF-Biot) equations and the Biot-Johnson-Koplik-Dashen (Biot-JKD) models are considered. In LF-Biot equations, the dissipation terms are proportional to the relative velocity between the fluid and the solid by a constant. Contrast to this, the dissipation terms in the Biot-JKD model are in the form of time convolution (memory) as a result of the frequency-dependence of fluid-solid interaction at the underlying microscopic scale in the frequency domain. The dynamic tortuosity and permeability described by Darcy's law are two crucial factors in this problem, and highly linked to the viscous force. In the Biot model, the key difficulty is to handle the viscous term when the  pore fluid is viscous flow. In the Biot-JKD dynamic permeability model, the convolution operator involves order $1/2$ shifted fractional derivatives in the time domain, which is challenging to discretize.

In this work, we utilize the  multipoint Pad$\acute{e}$  (or Rational) approximation for Stieltjes function to approximate the dynamic tortuosity and then obtain an augmented system of equations which avoids storing the solutions of the past time. The Runge-Kutta discontinuous Galerkin (RKDG) method is used to  compute the numerical solution, and numerical examples are presented to demonstrate the high order accuracy and stability of the method.

\end{abstract}


\end{frontmatter}

\section{Introduction}
\label{sec:1}
Poroelastic materials are abundant in nature, and wave propagation in those media is widely studied in many fields such as industrial foams, spongy bones, and petroleum rocks.  M. A. Biot  proposed the most widely used constitutive equations for wave propagation in a saturated porous medium, see  \cite{Biotlw, Biothf} in 1956 and \cite{Biot} in 1962.  Biot's theory predicts there are three waves propagating in isotropic poroelastic materials: fast P waves, which are analogous to standard elastic P wave, shear waves analogous to S waves, and slow P waves where the fluid expands while the solid contract, or vice versa.

For Biot's equations, there are two frequency regimes separated by the critical frequency $f_c$ which depends on the material properties; frequencies higher than $f_c$ are referred to as high frequency, otherwise as low frequency. In low frequency, the fluid within the pores is assumed to be of Poiseuille type, and the friction is a linear function of the relative velocity  between pore fluid and the solid matrix frame. The energy dissipates as a result of the viscous nature of the pore fluid and its interaction with the solid frame at the pore scale. The fast P waves and shear waves are lightly damped, and the slow P waves are strongly damped and propagate at a significantly slower speed than the other two waves through the medium. Therefore, the slow P waves are significant near the locations of the sources or material heterogeneities. The stiffness of the Biot system remains the major challenge in the simulation procedure. In high frequency, the viscous effect is proportional to the relative velocity by a constant that is frequency-dependent; this frequency dependence is to correct for the departure from laminar flow. In the time domain, this frequency dependence results in an order $1/2$ shifted fractional derivative. Therefore, modeling task in high frequency will face more challenges with discretizing the fractional derivative and storing the solutions of whole past steps. In 1956, Biot firstly gave an expression of the control mode in high frequency, which is effective for particular geometries \cite{Biothf}. In 1987, Johnson-Koplik-Dashen (JKD) \cite{JDK} put forward a general model of dynamic permeability, where the inertia interaction and viscous efforts are accounted for all frequencies. In this model, in the time domain, the fractional derivative which equals a time convolution product with a singular kernel was introduced. In 1993, \cite{Avellaneda}, Avellaneda and Torquato derived that the permeability can be represented as the Stieltjes function, which is valid for all pore space geometry. In 1993, Pride et al. derived a drag force model for more general materials geometries \cite{Pride2}.

Many researchers did a variety of works to describe poroelasticity analytically and numerically. Selvadurai  \cite{Selvadurai} gave a review of analytical solution for geomechanics. The uniqueness of the solutions for Biot's equations with the corresponding boundary conditions were studied by Deresiewicz and Skalak \cite{Deresiewicz}.  Detournay and Cheng \cite{Detournay} obtained an analytical solution in the Laplace transform domain and then transformed to the time domain.  Gilbert, Hsiao and Xu \cite{Gilbert} presented a variational formulation for the fluid-bone interaction problem by utilizing the modified Biot's equations and the boundary integral equation. Buchanan, Gilbert and Khashanah \cite{Buchanan} obtained the time-harmonic solutions for cancellous bone by using the finite element method. Numerical approaches for poroelasticity started from 1980s. Carcione et al. \cite{carcione review} provided a literature review for previous works. Garg \cite{Garg} used the finite difference method to model poroelasticity in one dimension, which seems to be the earliest numerical results, further works include Mikhailenko \cite{Mikhailenko}, Hassanzadeh \cite{Hassanzadeh}, Dai et al. \cite{Dai}, Chiavassa and Lombard \cite{Chiavassa}. Santos and Ore$\tilde{n}$a \cite{Santos} firstly used the finite element method to investigate the low frequency Biot's equations. Furthermore, other methods, such as the boundary element method by Attenborough, Berry, and Chen \cite{Attenborough}, the spectral element method \cite{Morency} and \cite{Sidler}  have been used to study this problem. Discontinuous Galerkin (DG) method also has been used to study poroelasticity, such as that of de la Puente et al. \cite{Puente} as well as the recent works of Chen, Luo and Feng \cite{Chen}, Dudley Ward, L$\ddot{a}$hivaara and Eveson \cite{Dudley Ward}.  Lemoine, Ou and Leveque \cite{Yvonne} firstly used high-resolution finite volume method to model poroelasticity in the time domain, and further works on the interaction of poroelastic and fluid were in \cite{Lemoine2}. Masson and Pride \cite{Masson} defined a time convolution product to discretize the fractional derivative. Lu and Hanyga \cite{Lu} developed a new method to calculate the shifted fractional derivative without storing and integrating the entire velocity histories. Recent years, Chiavassa and Lombard et al. \cite{Blance, Lombard2} used an optimization procedure to approach the fractional derivative.  Numerical works in the 1970s and 1980s focused on isotropic poroelasticity, and the earliest work on anisotropic poroelasticity was  Carcione \cite{carcione2} in 1996, where an algorithm for heterogeneous linear anisotropic porous media were developed for both low and high frequencies.

The purpose of this paper is to simulate wave propagation for transverse porous media. For the  {\blue LF-}Biot equations, we use the direct method to handle the viscous term, which is different from the splitting method \cite{Puente, Yvonne, Chiavassa}; both the single material and heterogeneous materials with point source tests are considered. For the Biot-JKD equations, we applied the technique developed in \cite{Ou-Woer} for dealing with the memory term and solve the augmented Biot-JKD equations, which deal with the memory terms by using auxiliary variables derived from the JKD-tortuosity.  The main idea in \cite{Ou-Woer} is to utilize the Stieltjes function representation of the JKD dynamic tortuosity in frequency domain \cite{YvonneTw} to obtain its multipoint Pad$\acute{e}$ approximation.  Analytical Inverse Laplace transform is then applied to this approximation to obtain the time-domain approximation of the memory term as a sum of relaxation kernel with different relaxation times  convoluted with the unknown relative velocity between fluid and solid. Each of these terms is referred to as an auxiliary memory variable and it can be easily shown to satisfy a local time ordinary differential equations, which is different from the models in \cite{Lu,Blance, Lombard2}. This new system, termed Augmented Biot-JKD equation in \cite{Ou-Woer} has no explicit memory terms, thus the procedure for handling the system is much easier due to the absence of the fractional derivative. For numerical simulations, we use the Runge-Kutta discontinuous Galerkin (RKDG) method. The first DG method was proposed in 1973 by Reed and Hill to solve a time independent linear  hyperbolic equation \cite{Reed}. The high order RKDG method was further developed by Cockburn and Shu \cite{Shu}, and thereafter, DG methods and their several variants have been successfully applied to study many problems in sciences and engineering.

This paper is organized as follows. In Section \ref{Gov}, we present the mathematical formulation of the governing equations. A brief description of the RKDG method is included in Section \ref{RKDG}. In Section \ref{Analytic}, we present the derivation of analytic plane waves solutions. Numerical examples are presented to illustrate the accuracy of the method and the properties of waves in Section \ref{numerics}. Finally, the conclusion is given in Section \ref{Conclude}.

\section{\label{Gov}Governing equations }
In this section, we will give the governing equations of wave propagation in poroelastic media.
The governing equations can be considered as the homogenized equations of the equations in the microscale (Stokes equations for the pore fluid and linear elastodynamics for the solid with no-slip conditions on the fluid-solid interface). Therefore, at the macroscopic level, the fluid and solid are considered to coexist at every point in the poroelastic material. Let $\mathbf{u}$ be the skeleton solid displacement vector, $\mathbf{U}$ the fluid displacement vector, and  $\mathbf{w}:=\phi(\mathbf{U}-\mathbf{u})$ be the relative motion of the fluid scaled by the porosity $\phi$ and $\zeta:=-\nabla\cdot\mathbf{w}$ be the variation in fluid content.  The  solid velocity $\mathbf{v}$ and the relative fluid velocity $\mathbf{q}$ are given as follows
\begin{equation}
\label{Eq:velocity}
\mathbf{v}:=\mathbf{\dot u} , \ \mathbf{q}:=\mathbf{\dot w},
\end{equation}
where the dot $\dot{}$ represents derivative with respect to time $t$. For the poroelastic problem, the critical frequency (Biot characteristic frequency) is defined by $f_{ci}:=\frac{\phi \eta}{2\pi\alpha_{\infty i}\kappa_{i}\rho_{f}}=\frac{\omega_{ci}}{2\pi},~i=x,y,z$, where $\eta$ is the dynamic viscosity of the pore fluid, $\rho_{f}$ the density of pore fluid, $\alpha_{\infty i}$ the static tortuosity and $\kappa_i$ the static permeability in the $x_i$ direction. Frequency below the critical frequency is low frequency regime, otherwise is high frequency regime.

\label{sec:2}
\subsection{Low frequency Biot Equations}
In this section, we present the wave equations in low frequency (Biot model). These contents are mainly based on \cite{Biotlw, Carcione, Yvonne}.

\subsubsection{The constitutive relation}
To describe Biot equation, we assume the solid matrix consists of isotropic material, and the pore space is connected and  filled with fluid. We also assume that the anisotropic of the solid matrix is resulted from the microstructure. Furthermore, the poroelastic materials are assumed to be transversal isotropic media, which have the properties of possessing three orthogonal symmetry axes. Let the $z$ axis be the symmetry axis, the stress-strain relations for a transversely isotropic elastic material has the following form:
\begin{equation}
\label{Eq:stree-strain}
\mbox{\boldmath{$\tau$}} = \mathbf{C} \mbox{\boldmath{$\epsilon$}}
\end{equation}
where the elastic stiffness tensor $\mathbf{C}$, stress variables \mbox{\boldmath{$\tau$}}, and strain variables \mbox{\boldmath{$\epsilon$}} are defined as
\begin{eqnarray}
\label{stress-strain}
  C&=&\left[
\begin{array}{cccccccc}
   c_{11}&c_{12}&c_{13}&0&0&0
 \\c_{12}&c_{11}&c_{13}&0&0&0
 \\c_{13}&c_{13}&c_{33}&0&0&0
 \\0&0&0&c_{55}&0&0
 \\0&0&0&0&c_{55}&0
 \\0&0&0&0&0&\frac{c_{11}-c_{12}}{2} \\
\end{array}
\right],\\
\mbox{\boldmath{$\tau$}}&=&(\tau_{11}\ \tau_{22}\ \tau_{33}\  \tau_{23}\  \tau_{13}\  \tau_{12}\ )^T,  \\
\mbox{\boldmath{$\epsilon$}}&=&(\epsilon_{11}\ \epsilon_{22}\  \epsilon_{33}\  2\epsilon_{23}\  2\epsilon_{13}\  2\epsilon_{12})^T,
\end{eqnarray}
with $\epsilon_{ij}=\frac{1}{2}\left(\partial_{i}u_{j}+\partial_{j}u_{i}\right)$.

\subsubsection{Strain energy for transversely isotropic poroelastic materials }
With the $z$ axis being the symmetry axis, the strain energy density of the Biot equation for transversal isotropic materials in three dimensional is defined as follows
\begin{equation}
\label{Eq:strainenergy}
\begin{aligned}
2V=&c^{u}_{11}(e^{(m)^2}_{11}+e^{(m)^2}_{22})+c^{u}_{33}e^{(m)^2}_{33}
+2c^{u}_{12}e^{(m)}_{11}e^{(m)}_{22}+2c^{u}_{13}\left(e^{(m)}_{11}+e^{(m)}_{22}\right)e^{(m)}_{33}\\
&+c^{u}_{55}\left(e^{(m)^2}_{23}+e^{(m)^2}_{13}\right)+c^{u}_{66}e^{(m)^2}_{12}
-2\beta_{1}M\left(e^{(m)}_{11}+e^{(m)}_{22}\right)\zeta-2\beta_{3}Me^{(m)}_{33}\zeta+M\zeta^2,
\end{aligned}
\end{equation}
where $e_{ii}^{(m)}:=\partial_{i}u_{i}$ and $e_{ij}^{(m)}:=\partial_{i}u_{j}+\partial_{j}u_{i},~ i\neq j$ are the strain components of the skeleton, $c_{ij}^u \in \mathbf{C}^u$ is the undrained elastic constants and $c^u_{66}=\frac{1}{2}(c^u_{11}-c^u_{12})$. The superscript $m$ represents the dry matrix and there exists relations:
\begin{eqnarray*}
\label{coefficients}
  c^u_{ij}&=& c^{(m)}_{ij}+M\beta_{i}\beta_{j}, \ \ i,j=1,...5,\\
  \mathbf{\beta}:&=& (\beta_{1}, ~\beta_{1}, ~\beta_{3}, ~0, ~0, ~0), \\
  \beta_{1}: &=& 1-\frac{c^{(m)}_{11}+c^{(m)}_{12}+c^{(m)}_{13}}{3K_{s}}, \\
  \beta_{3}: &=& 1-\frac{2c^{(m)}_{13}+c^{(m)}_{33}}{3K_{s}}, \\
  M: &=& \frac{K^2_{s}}{K_{s}\left[1+\phi(K_{s}/K_{f}-1)\right]-\left(2c^{(m)}_{11}+c^{(m)}_{33}
+2c^{(m)}_{12}+4c^{(m)}_{13}\right)/9},
\end{eqnarray*}
where $K_{s}$ and $K_{f}$ are the bulk modulus of the skeleton and pore fluid respectively. Having obtained the strain-energy expression, the total stresses (skeleton plus pore pressure) are given by:
\begin{equation}
\tau_{ij}=\frac{\partial V}{\partial e^{(m)}_{ij}},
\end{equation}
or equivalently
\begin{equation}
\label{Eq:porostress}
\tau_{I}=\sum_{J=1}^{6}c_{IJ}^u e_{J}^{(m)}-M \beta_{I}\zeta, \ \ I=1,...,6.
\end{equation}
Similarly, the formulation of pore pressure
\begin{equation}
\label{Eq:poropressure}
p=\frac{\partial V}{\partial \zeta}=M\left(\zeta-\sum\limits_{j=1}^{3}\beta_{j}e^{(m)}_{jj}\right).
\end{equation}

\subsubsection{Kinetic energy and dissipation potential}
The kinetic energy density in anisotropic medium can be expressed as
\begin{equation}
\label{Eq:kineticenergy}
T=\frac{1}{2}(\mathbf{\dot u}^T \mathbf{P}\mathbf{\dot u}
+2\mathbf{\dot u}^T \mathbf{R}\mathbf{\dot U}
+\mathbf{\dot U}^T \mathbf{T}\mathbf{\dot U}),
\end{equation}
where $\mathbf{R}$ is the induced mass matrix. Suppose all the three matrices have the diagonal form of $\mathbf{P}=\text{diag}(a_{1}, ~a_{2}, ~a_{3})$, $\mathbf{R}=\text{diag}(r_{1}, ~r_{2}, ~r_{3})$ and $\mathbf{T}=\text{diag}(t_{1}, ~t_{2}, ~t_{3})$. The mass coefficients has the significance that the relative fluid flow through the pores is not uniform and satisfy the equations
\begin{equation}
\label{masscoefficients}
a_{i}+r_{i}=(1-\phi)\rho_{s},  \ \ r_{i}+t_{i}=\phi\rho_{f},~i=1,2,3,
\end{equation}
where $\rho_{s}$ is the density of constituent solid, $r_{i}=\phi \rho_{f}(1-\alpha_{\infty i})$  is the induced parameter and $\rho=(1-\phi)\rho_{s}+\phi \rho_{f}$ is the bulk density of the media. The expression of (\ref{masscoefficients}) is based on the assumption that there is no relative motion between fluid and solid, more details can be found in \cite{Carcione}.

Based on the velocity vectors (\ref{Eq:velocity}), the kinetic energy density (\ref{Eq:kineticenergy}) can be rewritten as
\begin{equation}
\label{Eq:kineticenergy1}
\begin{aligned}
T&=\frac{1}{2}\sum_{i=1}^3\left[(1-\phi)\rho_{s}v_{i}^2-r_{i}\left(v_{i}-\dot U\right)^2+\phi \rho_{f} \dot U_{i}^2\right]\\
 &=\frac{1}{2}\sum_{i=1}^3\left[\rho v_{i}^2+2\rho_{f}q_{i}v_{i}+\left(\frac{\rho_{f}\phi-r_{i}}{\phi^2}\right)q_{i}^2\right]
   =:\tilde T(\mathbf{v},\mathbf{q}).
\end{aligned}
\end{equation}
In low frequency, the fluid flow inside the porous is assumed to be the Poiseuille type. In this situation, the dispassion potential in an anisotropic medium only depends on the relative motion between fluid and solid and has the form of
\begin{equation}
\Phi_{D}=\frac{\eta}{2} \mathbf{q}^T \mathbf{K}^{-1} \mathbf{q},
\end{equation}
where $\mathbf{K}$ is the {\blue static} permeability tensor.  Assume  $\mathbf{K}$ has the same principal direction as $\mathbf{P}$, $\mathbf{R}$, $\mathbf{T}$,  which indicates
$\mathbf{K}=\text{diag}(\kappa_{1}, ~\kappa_{2}, ~\kappa_{3})$.
This assumption simplifies the dissipation to
\begin{equation}
\Phi_{D}=\frac{1}{2}\sum_{i=1}^3\frac{\phi^2\eta}{\kappa_{i}}\left(\dot U_{i}-v_{i}\right)^2
        =\frac{1}{2}\sum_{i=1}^3\frac{\eta}{\kappa_{i}}q_{i}^2=: \tilde \Phi_{D}(\mathbf{q}).
\end{equation}

\subsubsection{Governing equations for plane-strain case}
In this work, we mainly consider the two dimensional plane-strain conditions in the $xz$ plane, thus all the components related to $e_{12}^{(m)}$, $e_{22}^{(m)}$ and $e_{23}^{(m)}$ will be omitted. Here we will present the governing equations of wave propagation in porous media. With the kinetic energy density and dissipation potential, the equations of motion for solid are given by
\begin{equation}
\partial_{t}\left(\frac{\partial T}{\partial v_{i}}\right)+\frac{\partial \Phi_{D}}{\partial v_{i}}=\sum_{j=1,3}\partial_{j}\left(\frac{\partial V}{\partial e_{ij}^{(m)}}+\phi p \delta_{ij}\right), ~i=1, 3,
\end{equation}
which is equivalent to
\begin{equation}
\label{eq:solidmotion}
\partial_{t}\left(\frac{\partial \tilde T}{\partial v_{i}}-\phi \frac{\partial \tilde T}{\partial q_{i}}\right)-\phi \frac{\partial \tilde \Phi_{D}}{\partial q_{i}}=\sum_{j=1,3}\partial_{j}(\frac{\partial V}{\partial e_{ij}^{(m)}}+\phi p \delta_{ij}).
\end{equation}
Similarly, the equations of motion for the fluid are
\begin{equation}
\partial_{t}\left(\frac{\partial T}{\partial \dot U_{i}}\right)+\frac{\partial \Phi_{D}}{\partial \dot U_{i}}=-\phi \sum_{j=1,3}\partial_{j}\left(\frac{\partial V}{\partial \zeta} \delta_{ij}\right), ~i=1, 3,
\end{equation}
or equivalently
\begin{equation}
\label{eq:fluidmotion}
\phi \partial_{t}\left(\frac{\partial \tilde T}{\partial q_{i}}\right)+\phi \frac{\partial \tilde \Phi_{D}}{\partial q_{i}}=-\phi \sum_{j=1,3}\partial_{j}\left(\frac{\partial V}{\partial \zeta}\delta_{ij}\right).
\end{equation}
The solid-fluid motion equations are given by adding (\ref{eq:solidmotion}) and (\ref{eq:fluidmotion})
\begin{equation}
\label{eq:solid-fluidmotion}
\sum\limits_{j=1,3}\partial _{j}\tau_{ij}=\rho \dot{v}_{i}+\rho_{f}\dot{q}_{i},~i=1, 3.
\end{equation}
Since $p=\frac{\partial V}{\partial \zeta}$, equation (\ref{eq:fluidmotion}) can be reduced to
\begin{equation}
\label{Eq:darcylow}
-\partial_{i}p=\rho_{f}\dot{v}_{i}+(\frac{\rho_{f}\phi-r_{i}}{\phi^2})\dot{q}_{i}+\frac{\eta}{\kappa_{i}}{q}_{i}= \rho_{f}\dot{v}_{i}+(\frac{\rho_{f}}{\phi})\alpha_{\infty i} \dot{q}_{i}+\frac{\eta}{\kappa_{i}} {q}_{i}, ~i=1, 3.
\end{equation}
The motion equation (\ref{eq:solid-fluidmotion}) is known as the conservation of momentum equations, and (\ref{Eq:darcylow}) is the Darcy's law in the low frequency regime. The two equations can be rewritten as the following explicit form of
\begin{eqnarray}
\label{Eq:motions}
  \partial_{t}v_{x} &=& \frac{1}{\Delta_{1}}\left(m_{1}\partial_{x}\tau_{xx}+
m_{1}\partial_{z}\tau_{xz}+\rho_{f}\partial_{x}p+\rho_{f}\frac{\eta}{\kappa_{1}}q_{x}\right), \\
  \partial_{t}q_{x} &=& \frac{1}{\Delta_{1}}\left(-\rho_{f}\partial_{x}\tau_{xx}-
\rho_{f}\partial_{z}\tau_{xz}-\rho\partial_{x}p-\rho\frac{\eta}{\kappa_{1}}q_{x}\right),\\
  \partial_{t}v_{z} &=& \frac{1}{\Delta_{3}}\left(m_{3}\partial_{x}\tau_{xz}+
m_{3}\partial_{z}\tau_{zz}+\rho_{f}\partial_{z}p+\rho_{f}\frac{\eta}{\kappa_{3}}q_{z}\right), \\
\label{Eq:motions1}
  \partial_{t}q_{z} &=& \frac{1}{\Delta_{3}}\left(-\rho_{f}\partial_{x}\tau_{xz}-
\rho_{f}\partial_{z}\tau_{zz}-\rho\partial_{z}p-\rho\frac{\eta}{\kappa_{3}}q_{z}\right),
\end{eqnarray}
where $m_{i}:=\frac{\rho_{f}\phi-r_{i}}{\phi^2}=\frac{\rho_{f}\alpha_{\infty i}}{\phi}$,~$\Delta_{i}:=\rho m_{i}-\rho_{f}^2,~i=1,3$ and the subsccript  $j=1 \rightarrow j=x$,~ $j=3 \rightarrow j=z$.\\

Furthermore, assume the solid and fluid external sources are $\mathbf{S}=(s_1,~s_3,~s_5,~0,~0,~s_f,~0,~0)^T$, by differentiating  equations (\ref{Eq:porostress}) and (\ref{Eq:poropressure}) with respect to time, we can gain the stress-strain relations for the plane-strain case:
\begin{eqnarray}
\label{Eq:plane-strain}
  \partial_{t}\tau_{xx} &=& c_{11}^u \partial_{x}v_{x}+c_{13}^u \partial_{z}v_{z}
+\beta_{1}M(\partial_{x}q_{x}+\partial_{z}q_{z})+\partial_{t}s_{1}, \\
  \partial_{t}\tau_{zz} &=& c_{13}^u \partial_{x}v_{x}+c_{33}^u \partial_{z}v_{z}
+\beta_{3}M(\partial_{x}q_{x}+\partial_{z}q_{z})+\partial_{t}s_{3}, \\
  \partial_{t}\tau_{xz} &=& c_{55}^u (\partial_{z}v_{x}+\partial_{x}v_{z}) + \partial_t s_5, \\
  \label{Eq:plane-strain1}
  \partial_{t}p &=& -\beta_{1}M\partial_{x}v_{x}-\beta_{3}M\partial_{z}v_{z}
-M(\partial_{x}q_{x}+\partial_{z}q_{z})+\partial_{t}s_{f}.
\end{eqnarray}

The first order system for wave propagation in poroelastic media consists of the motion equations (\ref{Eq:motions})-(\ref{Eq:motions1}) and the stress-strain relations (\ref{Eq:plane-strain})-(\ref{Eq:plane-strain1})
in the following form of balance law
\begin{equation}
\label{eq:lowsyms}
\partial_{t}\mathbf{Q}+\triangledown \cdot \mathbf{F}(\mathbf{Q})
=\mathbf{h}(\mathbf{Q})+\partial_{t}\mathbf{S},
\end{equation}
where $\mathbf{Q}=(\tau_{xx}, ~\tau_{zz}, ~\tau_{xz}, ~v_{x}, ~v_{z}, ~p, ~q_{x}, ~q_{z})^T$ is the unknown vector, $\mathbf{F}=(\mathbf{f}, ~\mathbf{g})$ is the flux function and $\mathbf{h}(\mathbf{Q})$ is the viscous term, $\mathbf{S}$ is the external source, via
\begin{equation}
\begin{aligned}
\mathbf{f}(\mathbf{Q})=&-\left(c_{11}^u v_{x}+\beta_{1} M q_{x},~ c_{13}^u v_{x}+\beta_{3} M q_{x}, ~c_{55}^u v_{z},~ \frac{m_{1}}{\Delta_{1}}\tau_{xx}+\frac{\rho_{f}}{\Delta_{1}}p,~
\frac{m_{3}}{\Delta_{3}}\tau_{xz},~-\beta_{1}Mv_{x}-Mq_{x},~\right.\\
&\left.-\frac{\rho_{f}}{\Delta_{1}}\tau_{xx}-\frac{\rho}{\Delta_{1}}p,~-\frac{\rho_{f}}{\Delta_{3}}\tau_{xz}\right)^T,
\end{aligned}
\end{equation}
\begin{equation}
\begin{aligned}
\mathbf{g}(\mathbf{Q})=&-\left(c_{13}^u v_{z}+\beta_{1} M q_{z},~ c_{33}^u v_{z}+\beta_{3} M q_{z},~c_{55}^u v_{x},
~ \frac{m_{1}}{\Delta_{1}}\tau_{xz},~
\frac{m_{3}}{\Delta_{3}}\tau_{zz}+\frac{\rho_{f}}{\Delta_{3}}p,~
-\beta_{3}Mv_{z}-Mq_{z}\right.,\\
&\left.-\frac{\rho_{f}}{\Delta_{1}}\tau_{xz},~
-\frac{\rho_{f}}{\Delta_{3}}\tau_{zz}-\frac{\rho}{\Delta_{3}}p\right)^T,
\end{aligned}
\end{equation}
\begin{eqnarray}
  \mathbf{h}(\mathbf{Q})&=& \left(0, ~0, ~0, ~\frac{\rho_{f}\eta}{\Delta_{1}\kappa_{1}}q_{x},~
\frac{\rho_{f}\eta}{\Delta_{3}\kappa_{3}}q_{z},~0,~
-\frac{\rho \eta}{\Delta_{1}\kappa_{1}}q_{x},~
-\frac{\rho \eta}{\Delta_{3}\kappa_{3}}q_{z}\right)^T.
  \label{eq:flux}
\end{eqnarray}
It was proved that the system (\ref{eq:lowsyms}) is a hyperbolic system in \cite{Yvonne}.

\subsection{Biot-JKD equations}
In the principal coordinates, the general poroelastic wave  equations have the following form
\beqa
\sum_{k=1}^3 \frac{\partial \tau_{jk}}{\partial x_k}=\rho \frac{\partial v_j}{\partial t} + \rho_f \frac{\partial q_j}{\partial t},~\, t>0 , \label{test}\\
-\frac{\partial p}{\partial x_j} = \rho_f \frac{\partial v_j}{\partial t}+ \left(\frac{\rho_f}{\phi} \right) \check{\alpha_j} \star \frac{\pr q_j }{\pr t},~\,t>0,\, j=1,2,3,
\label{convo}
\eeqa
where $\star$ denotes the time-convolution operator and $\check{\alpha_j}$ is the inverse Laplace transform of the dynamic tortuosity $\alpha_j(\omega)$ with $\omega$ being the frequency. Here the one-sided  Laplace transform of a function $f(t)$ is defined as
\[
\hat{f}(\omega):={\cal{L}}[f](s=-i\omega):=\frac{1}{\sqrt{2\pi}} \int_{0}^{\infty} f(t)e^{-st} dt .
\]
We note that the viscodynbamic operator $Y$ defined in Equation (7.213) in \cite{Carcione} is related to $\check{\alpha_j}$ by a constant,
\[
Y=\left(\frac{\rho_f}{\phi} \right) \check{\alpha_j},\, j=1,2,3.
\]
As a special case of the general Biot equations, equation \eqref{Eq:darcylow}  in low frequency Biot's equations corresponds to
\[
\check{\alpha_j}(t)=\alpha_{\infty j} \delta(t)+\frac{\eta \phi}{\kappa_{j}\rho_f}H(t),\, j=1,2,3,
\]
where $\delta(t)$ is the Dirac function and $H(t)$ the Heaviside function. This low-frequency tortuosity function corresponds to
\[
\alpha_j(\omega)=\alpha_{\infty j}+\frac{\eta \phi/\kappa_{j} \rho_f}{-i\omega},\, j=1,2,3 .
\]
The Biot-JKD equations refer to a special case when  $\alpha_j$ in the general Biot's equations is  the JKD dynamic tortuosity function $T_j(\omega)$ defined in \cite{JDK}
\begin{equation}
\label{Eq:Tortuosity}
\alpha_j(\omega)=T_j(\omega)=\alpha_{\infty j}+\frac{i\eta\phi}{\omega \kappa_{j}\rho_f}\left(1-\frac{4i\alpha^2_{\infty j}\kappa^2_{j}\rho_f\omega}{\eta \Lambda_j^2\phi^2}\right)^{\frac{1}{2}},~ j=1,2,3.
\end{equation}

It is shown in \cite{YvonneTw} that $T_j(\omega)$ has a simple pole at $\omega=0$ with strength $\frac{\eta\phi}{\rho_f \kappa_{j}}$ and has a branch cut on the $\omega$-plane along the half line $-i \omega\in [-\infty,~-\frac{\nu \phi^2 \Lambda^2}{4 \alpha_{\infty j}^2 \kappa_{j}^2}]$ after choosing an appropriate branch. Utilizing the representation formula for $T_j(\omega)$ derived in \cite{YvonneTw}, it was proved in \cite{Ou-Woer} that for $\bq(\bx,t), \, t\le 0$, the Biot-JKD equations can be very well approximated by the following so-called Augmented Biot-JKD equations that have no explicit memory terms,

\begin{empheq}[left=\empheqlbrace]{align}
\sum_{k=1}^3 \frac{\partial \tau_{jk}}{\partial x_k}& =\rho \frac{\partial v_j}{\partial t} + \rho_f \frac{\partial q_j}{\partial t},\, t>0  ,
\label{aug_b}
\\
\pr_t\Theta_k^{x_j}(\bx,t)&=p_k^j \Theta_k^{x_j}(\bx,t)-p_k^j q_j(\bx,t),\, j=1,2,3,
\label{aug_theta}
\\
-\frac{\partial p}{\partial x_j} &= \rho_f \frac{\partial v_j}{\partial t}+ \left(\frac{\rho_f \alpha_{\infty j}}{\phi} \right) \frac{\pr q_j }{\pr t}
  + \left( \frac{\eta}{\kappa_{j}}+\frac{\rho_f}{\phi}\sum_{k=1}^M r_k^j \right) q_j \nonumber\\
 &-\left(\frac{\rho_f}{\phi}\right)\sum_{k=1}^M r_k^j \Theta_k^{x_j},\,t>0,\, j=1,2,3.
 \label{aug_e}
\end{empheq}
where the auxiliary variables $\Theta_k^{x_j}$ with $j=1,2,3$ and $k=1,\cdots,M$ are defined as
\beq
\Theta_k^{x_j}(\bx,t):= (-p_k) e^{p_k^j t} \star  q_j.
\label{theta_def}
\eeq
The constants $r_k^j$ and $p_k^j$, $k=1,\cdots,M$ are related to the the pole-residue approximation for $\alpha_j(\omega)$ for the frequency range relevant to the point source spectral content; they can be computed with very high accuracy with the algorithms presented in \cite{Ou-Woer}. The representation formula of $\alpha_j(\omega)$ guarantees that $r_k^j>0$ and $p_k^j<0$.
\vspace{0.1in}
With this notation, the Darcy's law for the plain strain case in the $xz$ plane can be written as ($j=1 \rightarrow j=x$, $j=3 \rightarrow j=z$)
\begin{equation}
\label{Eq:finalDarcy}
\begin{aligned}
-\nabla p &=\rho_f \partial_{t}\mathbf{v}+\frac{\rho_f}{\phi}\text{diag}(\alpha_{\infty x},~\alpha_{\infty z})\partial_{t}\mathbf{q}
+\text{diag}\left(\frac{\eta}{\kappa_x}+\frac{\rho_f}{\phi}\sum_{j=1}^M r^x_j,~\frac{\eta}{\kappa_z}+\frac{\rho_f}{\phi}\sum_{j=1}^M r^z_j\right)\mathbf{q}\\
&-\text{diag}\left(\frac{\rho_f}{\phi}\sum_{j=1}^{M}r^x_j\Theta^x_j(t),
~\frac{\rho_f}{\phi}\sum_{j=1}^{M}r^z_j\Theta^z_j(t)\right)\mathbf{I},
\end{aligned}
\end{equation}
where $\mathbf{I}$ is the $2 \times 2$ unit matrix. The motion equations of the fluid-solid are the same as the low frequency case
\begin{equation}
\label{Eq:highmotion}
\sum\limits_{j=x,z}\partial _{j}\tau_{ij}=\rho \dot{v}_{i}+\rho_{f}\dot{q}_{i},  \ \ i=x,z.
\end{equation}
Combining (\ref{Eq:finalDarcy}) and (\ref{Eq:highmotion}), we then get the following motion equations
\begin{equation}
\label{Eq:vxh}
\partial_t v_x=\frac{1}{\Delta_1}\left[m_1\partial_x\tau_{xx}+m_1\partial_z\tau_{xz}
+\rho_f\left(\frac{\eta}{\kappa_1}+\frac{\rho_f}{\phi}\sum_{j=1}^{M^x}r_j^x\right)q_x+\rho_f\partial_x p
-\frac{\rho_f^2}{\phi}\sum_{j=1}^{M_x}r_j^x\Theta_j^x(t)\right],
\end{equation}
\begin{equation}
\label{Eq:qxh}
\partial_t q_x=\frac{1}{\Delta_1}\left[-\rho_f\partial_x\tau_{xx}-\rho_f\partial_z\tau_{xz}
-\rho\left(\frac{\eta}{\kappa_1}+\frac{\rho_f}{\phi}\sum_{j=1}^{M_x}r_j^x\right)q_x-\rho\partial_x p
+\frac{\rho_f\rho}{\phi}\sum_{j=1}^{M_x}r_j^x\Theta_j^x(t)\right].
\end{equation}
Similarly
\begin{equation}
\label{Eq:vzh}
\partial_t v_z=\frac{1}{\Delta_3}\left[m_3\partial_x\tau_{xz}+m_3\partial_z\tau_{zz}
+\rho_f\left(\frac{\eta}{\kappa_3}+\frac{\rho_f}{\phi}\sum_{j=1}^{M_z}r_j^z\right)q_z+\rho_f\partial_z p
-\frac{\rho_f^2}{\phi}\sum_{j=1}^{M_z}r_j^z\Theta_j^z(t)\right],
\end{equation}
\begin{equation}
\label{Eq:qzh}
\partial_t q_z=\frac{1}{\Delta_3}\left[-\rho_f\partial_x\tau_{xz}-\rho_f\partial_z\tau_{zz}
-\rho\left(\frac{\eta}{\kappa_3}+\frac{\rho_f}{\phi}\sum_{j=1}^{M_z}r_j^z\right)q_z-\rho\partial_z p
+\frac{\rho_f\rho}{\phi}\sum_{j=1}^{M_z}r_j^z\Theta_j^z(t)\right].
\end{equation}
Furthermore, the stress-strain relations (\ref{Eq:plane-strain})-(\ref{Eq:plane-strain1}) stay the same as the low-frequency case
\begin{eqnarray}
\label{Eq:plane-strainh}
  \partial_{t}\tau_{xx} &=& c_{11}^u \partial_{x}v_{x}+c_{13}^u \partial_{z}v_{z}
+\beta_{1}M(\partial_{x}q_{x}+\partial_{z}q_{z})+\partial_{t}s_{1}, \\
  \partial_{t}\tau_{zz} &=& c_{13}^u \partial_{x}v_{x}+c_{33}^u \partial_{z}v_{z}
+\beta_{3}M(\partial_{x}q_{x}+\partial_{z}q_{z})+\partial_{t}s_{3}, \\
  \partial_{t}\tau_{xz} &=& c_{55}^u (\partial_{z}v_{x}+\partial_{x}v_{z}) + \partial_t s_5,\\
\label{Eq:plane-strainh1}
  \partial_{t}p &=& -\beta_{1}M\partial_{x}v_{x}-\beta_{3}M\partial_{z}v_{z}
-M(\partial_{x}q_{x}+\partial_{z}q_{z})+\partial_{t}s_{f}.
\end{eqnarray}
The governing equations (\ref{Eq:vxh})-(\ref{Eq:plane-strainh1}) can be written in the following form of balance law.
\begin{equation}
\label{eq:higheqn}
\partial_{t}\mathbf{Q}+\triangledown \cdot \mathbf{F}(\mathbf{Q})
=\mathbf{h}(\mathbf{Q})+\partial_{t}\mathbf{S},
\end{equation}
where $\mathbf{Q}=\left(\tau_{xx},~ \tau_{zz},~ \tau_{xz},~ v_{x},~ v_{z},~ p,~ q_{x},~ q_{z},~ \Theta^x_{1,...,M^x},~ \Theta^z_{1,...,M^z}\right)^T$ is the unknown vector,
$\mathbf{F}=(\mathbf{f},\mathbf{g})$ is the flux function, $\mathbf{h}(\mathbf{Q})$ is the viscous term, $\mathbf{S}$ is the external source, via
\begin{equation}
\label{eq:F fluxhigh}
\begin{aligned}
\mathbf{f}(\mathbf{Q})=&-\left(c_{11}^u v_{x}+\beta_{1} M q_{x},~ c_{13}^u v_{x}+\beta_{3} M q_{x},~c_{55}^u v_{z},~ \frac{m_{1}}{\Delta_{1}}\tau_{xx}+\frac{\rho_{f}}{\Delta_{1}}p,~
\frac{m_{3}}{\Delta_{3}}\tau_{xz},~-\beta_{1}Mv_{x}-Mq_{x},\right.\\
&\left.-\frac{\rho_{f}}{\Delta_{1}}\tau_{xx}-\frac{\rho}{\Delta_{1}}p,~
-\frac{\rho_{f}}{\Delta_{3}}\tau_{xz},~0_{1,...,M^x},~0_{1,...,M^z}\right)^T,
\end{aligned}
\end{equation}
\begin{equation}
\label{eq:G fluxhigh}
\begin{aligned}
\mathbf{g}(\mathbf{Q})=&-\left(c_{13}^u v_{z}+\beta_{1} M q_{z},~ c_{33}^u v_{z}+\beta_{3} M q_{z},~c_{55}^u v_{x}, ~ \frac{m_{1}}{\Delta_{1}}\tau_{xz},~
\frac{m_{3}}{\Delta_{3}}\tau_{zz}+\frac{\rho_{f}}{\Delta_{3}}p,~
-\beta_{3}Mv_{z}-Mq_{z},\right.\\
&\left.-\frac{\rho_{f}}{\Delta_{1}}\tau_{xz},~
-\frac{\rho_{f}}{\Delta_{3}}\tau_{zz}-\frac{\rho}{\Delta_{3}}p,~0_{1,...,M^x},~0_{1,...,M^z}\right)^T,
\end{aligned}
\end{equation}
\begin{equation}
\label{Eq:visfluxhigh}
\begin{aligned}
\mathbf{h}(\mathbf{Q})=&\left(0,~0,~0,~\left(\frac{\rho_{f}\eta}{\Delta_{1}k_{1}}+\frac{\rho_f^2}
{\Delta_1\phi}\sum_{j=1}^{M_x}r_j^x\right)q_{x}-\frac{\rho_f^2}{\Delta_1 \phi}\sum_{j=1}^{M^x}r_j^x\Theta_j^x,~ \
\left(\frac{\rho_{f}\eta}{\Delta_{3}k_{3}}+\frac{\rho_f^2}
{\Delta_3\phi}\sum_{j=1}^{M_z}r_j^z\right)q_{z}\right.\\
&-\frac{\rho_f^2}{\Delta_3 \phi}\sum_{j=1}^{M^z}r_j^z\Theta_j^z,~ \ 0,~
-\left(\frac{\rho \eta}{\Delta_{1}k_{1}}+\frac{\rho \rho_f}{\Delta_1\phi}\sum_{j=1}^{M^x}r_j^x\right)q_{x}+\frac{\rho\rho_f}{\Delta_1\phi}\sum_{j=1}^{M^x}r_j^x
\Theta_j^x,~\\
& ~-\left(\frac{\rho \eta}{\Delta_{3}k_{3}}+\frac{\rho \rho_f}{\Delta_3\phi}\sum_{j=1}^{M^z}r_j^z\right)q_{z}+\frac{\rho\rho_f}{\Delta_3\phi}\sum_{j=1}^{M^z}r_j^z
\Theta_j^z,~\\
&-p^x_{1}(q_x-\Theta^x_1),...,-p^x_{M^x}(q_x-\Theta^x_{M^x}),~-p^z_{1}(q_z-\Theta^z_1),
...,-p^z_{M^z}(q_z-\Theta^z_{M^z}) \Bigg)^T,
\end{aligned}
\end{equation}
\begin{equation}
\mathbf{S}=\left(s_1,~s_3,~s_5,~0,~0,~s_f,~0,~0,~0_{1,...,M^x},~0_{1,...,M^z}\right)^T.
\end{equation}
We note that the system (\ref{eq:higheqn}) - (\ref{Eq:visfluxhigh}) has the same structure as the one for LF-Biot equations, and call this system as the Augmented Biot-JKD equations since the auxiliary variables are added to the original system  yielding a new  system with a larger scale. The auxiliary variables in (\ref{theta_def}) satisfying a pure ODE system (\ref{aug_theta}) are different with those in \cite{Lombard2, Lu, Dudley Ward}. In the Augmented Biot-JKD equations, the auxiliary variables only depend on  themselves and the relative fluid velocities, and thus the related components of the flux functions are all zero, making the system very simple to be simulated for interface fluxes when heterogeneous materials are considered.

All our numerical simulations are based on equations (\ref{eq:lowsyms}) and (\ref{eq:higheqn}). In low frequency, the system (\ref{eq:lowsyms}) has been proved to be a hyperbolic system in \cite{Yvonne}.  In high frequency, according to the property of auxiliary variables, we claim that the system (\ref{eq:higheqn}) is also a hyperbolic system because the structure of its Jacobian is the same as low frequency case except for zero components.

\section{\label{RKDG}RKDG method}
In this section, we apply the Runge-Kutta discontinuous Galerkin (RKDG) method \cite{Shu} to explore the first order hyperbolic systems (\ref{eq:lowsyms}) and (\ref{eq:higheqn}).
\subsection{Space discretization}
Let the computational domain be $\Omega=[b_1,~b_2]\times [c_1,~c_2]$ and be partitioned by an uniform rectangular mesh $I=\bigcup I_{i,j}, ~ i=1,...N_{x}, ~j=1,...N_{z}$, where $N_{x},~N_{z}$ are the total number of cells in $x$ axis and $z$ axis respectively and $I_{i,j}=[x_{i-1/2},~ x_{i+1/2}]\times [z_{i-1/2}, ~z_{i+1/2}]$ for $1\leq i \leq N_{x}$ and $1\leq j \leq N_{z}$.  Here, $b_1=x_{1/2}<x_{3/2}<...<x_{N_{x}+1/2}=b_2 $ and $ c_1=z_{1/2}<z_{3/2}<...<z_{N_{z}+1/2}=c_2$.

Then, we define the approximation space consisting of piecewise polynomials
\begin{equation}
V_{h}^n=\left\{v:v|_{I_{i,j}}\in P^n(I_{i,j}),\,\, i=1,...N_{x},\, j=1,... N_{z}\right\},
\end{equation}
where $P^n(I_{i,j}) $ denotes the set of polynomials of degree up to $n$ defined on $I_{i,j}$.  For instance, choosing $\varphi _{i,j}^l(x,z),~ l=1,2,...,deg=(n+1)(n+2)/2$ to be the local basis of $P^n(I_{i,j})$, the numerical solution, $\mathbf{Q}_{h}$, inside $I_{i,j}$ then can be expressed as
\begin{equation}
\mathbf{Q}_{h}(x,z,t)=\sum_{l=1}^{deg}q_{i,j}^l(t) \varphi_{i,j}^l(x,z).
\end{equation}

In order to seek the approximate solution $\mathbf{Q}_{h}\in V_{h}^n$, we consider the weak formulations of (\ref{eq:lowsyms}) and (\ref{eq:higheqn}).  Multiplying systems (\ref{eq:lowsyms}) and (\ref{eq:higheqn}) with a test function $v(x,z)$, and integrating by parts over a cell $I_{i,j}$, we obtain
\begin{equation}
\begin{aligned}
\frac{d}{dt}\int_{I_{i,j}}\mathbf{Q}(x,z,t)v(x,z)dxdz
+\sum_{e\in \partial I_{i,j}}\int_{e}
\mathbf{F}(\mathbf{Q}(x,z,t))\cdot \mathbf{n}_{e,I_{i,j}} v(x,z)ds \\
- \int_{I_{i,j}}\mathbf{F}(\mathbf{Q}(x,z,t)\nabla v(x,z)dxdz
=\int_{I_{i,j}}\mathbf{h}(\mathbf{Q}(x,z,t)v(x,z)dxdz,
\end{aligned}
\end{equation}
where $ds$ is the boundary measure and $\mathbf{n}_{e}=(n_{x},~n_{z})$ is the outward unit normal to edge $e$ of the cell $I_{i,j}$. Now assume both the solution $\mathbf{Q}$ and the test function $v$ comes from the approximation space $V_h^n$, the DG scheme is:

Find $\mathbf{Q}_h \in V_h^n$ such that
\begin{equation}
\begin{aligned}
&\frac{d}{dt}\int_{I_{i,j}}\mathbf{Q}_{h}(x,z,t)v_{h}(x,z)\\
&+\sum_{e\in \partial I_{i,j}}\int_{e}h_{e,I_{i,j}}\left(\mathbf{Q}_{h}\left(x^{int(I_{i,j})},z^{int(I_{i,j})},t\right),
\mathbf{Q}_{h}\left(x^{ext(I_{i,j})},z^{ext(I_{i,j})},t\right)\right)v_{h}(x,z)ds\\
&- \int_{I_{i,j}}\mathbf{F}(\mathbf{Q}_{h}(x,z,t)\nabla v_h(x,z)dxdz
=\int_{I_{i,j}}\mathbf{h}(\mathbf{Q}_{h}(x,z,t)v_h(x,z)dxdz,\,\, \text{~for all~} v_h(x,z) \in V_h,
\end{aligned}
\end{equation}
where $h_{e,I_{i,j}}(x,z,t)$ is the numerical flux. In this work, we use the Lax-Friedrichs flux (local)
\begin{equation}
h_{e,I_{i,j}}(a,b)=\frac{1}{2}\left[\mathbf{F}(a)\cdot \mathbf{n}_{e,I_{i,j}}+\mathbf{F}(b)\cdot \mathbf{n}_{e,I_{i,j}}-\alpha_{e,I_{i,j}}(b-a)\right],
\end{equation}
where $\mathbf{F}$ is the flux function and $\alpha_{e,I_{i,j}}$ is an estimate of the biggest eigenvalue of Jacobian matrix in a neighborhood of the edge $e$ of the cell $I_{i,j}$.

\subsection{Time discretization}
After the space discretization, we can obtain an ODE system of $\frac{dq(t)}{d t}=L_h(q(t))$ by inverting the mass matrix.
For the fully discrete system, we use strong stability preserving (SSP) high order Runge-Kutta \cite{shutime} time discretizations to achieve better accuracy in time. Such discretizations can be expressed as a convex combination of the forward Euler method, and thus they maintain strong stability properties in any semi-norm  of the forward Euler step. Let $\{t^n\}_{n=0}^N$ be a partition of $[0, \,T]$, with time step $\Delta t$, the third order total variation diminishing (TVD) Runge-Kutta method is given as follows:
\begin{eqnarray}
\label{Eq:time}
q^{(1)} &=& q^n + \Delta t L(q^n, t^n), \\
q^{(2)} &=& \frac{3}{4} q^n+ \frac{1}{4} q^{(1)}+ \frac{1}{4} \Delta t L(q^{(1)}, t^n+\Delta t), \\
q^{n+1} &=& \frac{1}{3} q^n+ \frac{2}{3} q^{(2)}+ \frac{2}{3} \Delta t L(q^{(2)}, t^n+\frac{1}{2}\Delta t).
\end{eqnarray}
Also, it is known that the CFL number is very important for the stability of the method.  The time step $\Delta t$ is dynamically determined by
\begin{equation}
\Delta t=\frac{\text{CFL}}{\frac{\max(|s_x|)}{\Delta x}+\frac{\max(|s_z|)}{\Delta z}},
\end{equation}
where $s_x$ and $s_z$ are the speeds in $x$ and $z$ directions respectively, the maxima are computed over all cell elements.

\subsection{Limiter, boundary and initial conditions}
In the simulation, if the solutions are not smooth and the RKDG method is applied, limiting operators are often needed to enhance the stability of the method and eliminate spurious oscillations in the numerical solutions. In this work, the total variation bounded (TVB) minmod slope limiter is applied to the numerical examples in Sections \ref{sec_plane_wave}  to \ref{sec_couple_JKD}. The limiters can be implemented either component by component or in local characteristic fields. Usually, the limiter can improve the behavior of the solution, but it also decreases the order of accuracy of the solutions.

In this work, we deal with the boundary conditions by using ghost cells. For the plane wave test cases, the boundary conditions are implemented through the $L^2$ projection of the exact boundary data onto the ghost cells. Outflow boundary conditions are used in all the exterior source examples, where the ghost elements values are treated by using the same polynomial representation as in the adjacent interior elements. The numerical solutions are initialized in the same way as boundary conditions when investigating the convergence results and are set to be $\mathbf{0}$ in the point source examples.

\section{\label{Analytic}Analytic plane wave solution}
This section is mainly about the analytic solution for the first order equation.  We use the same method given in \cite{Yvonne} to determine the analytic solution. Assume that the particle velocity and stress have the plane wave form, we define the velocity vector
\begin{equation*}
\mathbf{V}=(v_{x}, ~v_{z}, ~q_{x}, ~q_{z})^T=\mathbf{V}_{0} \exp(i(k_{x}x+k_{z}z-\omega t)),
\end{equation*}
and the stress vector
\begin{equation*}
\mathbf{T}=(\tau_{xx}, ~\tau_{zz}, ~\tau_{xz},~-p)^T=\mathbf{T}_{0} \exp(i(k_{x}x+k_{z}z-\omega t)),
\end{equation*}
where $\mathbf{V}_{0}, ~\mathbf{T}_{0}$ are constant vectors which relate to the amplitudes, $\omega$ is the prescribed angular frequency.
$\vec{k}=(k_{x},~k_{z})=(kl_{x},~kl_{z})$ is the wave vector to be determined, $k$ is the wave number and $l_{x}$ and $l_{z}$ are the direction cosine satisfying $l_{x}^2+l_{z}^2=1$. Substituting the plane wave to the stress-strain equations (\ref{Eq:plane-strain})-(\ref{Eq:plane-strain1}) or (\ref{Eq:plane-strainh})-(\ref{Eq:plane-strainh1}) yields
\begin{equation}
\label{Eq:planewave1}
-\omega\mathbf{T}_{0}=k\mathbf{F}\mathbf{V}_{0},
\end{equation}
where $\mathbf{F}$ is a matrix with the form of
\begin{equation*}
 \mathbf{F}=
\left[
\begin{array}{cccc}
   l_{x}c_{11}^u&l_{z}c_{13}^u&\beta_{1}Ml_{x}&\beta_{1}Ml_{z}
 \\l_{x}c_{13}^u&l_{z}c_{33}^u&\beta_{3}Ml_{x}&\beta_{3}Ml_{z}
 \\l_{z}c_{55}^u&l_{x}c_{55}^u&0&0
 \\ \beta_{1}Ml_{x}&\beta_{3}Ml_{x}&Ml_{x}&Ml_{z}
\end{array}
\right].
\end{equation*}
Substituting the plane wave to the motion equations (\ref{eq:solid-fluidmotion})-(\ref{Eq:darcylow}) and (\ref{Eq:finalDarcy})-(\ref{Eq:highmotion}) gives another system as
\begin{equation}
\label{Eq:planewave2}
k\mathbf{L}\mathbf{T}_{0}=-\omega\mathbf{\Gamma}\mathbf{V}_{0},
\end{equation}
where $\mathbf{\Gamma}$ and $\mathbf{L}$ are matrices, given by
\begin{eqnarray*}
  \mathbf{L} &=& \left[
\begin{array}{cccc}
   l_{x}&0&l_{z}&0
 \\0&l_{z}&l_{x}&0
 \\0&0&0&l_{x}
 \\0&0&0&l_{z}
\end{array}
\right],\\
  \mathbf{\Gamma} &=& \left[
\begin{array}{cccc}
   \rho & 0 & \rho_{f} & 0
 \\0 & \rho & 0& \rho_{f}
 \\\rho_{f} & 0 & -iY_{1}(\omega)/\omega & 0
 \\0 & \rho_{f} & 0 & -iY_{3}(\omega)/\omega
\end{array}
\right].
\end{eqnarray*}
For the low frequency Biot equations, we have $Y_{j}(\omega):=i\omega m_{j}-\eta/\kappa_{j}$ for j=1,~3; for the augmented Biot-JKD equations, $Y_{j}(\omega):=i\omega m_j-\frac{\eta}{\kappa_j}+i\omega\frac{\rho_f}{\phi}\sum_{l=1}^M\frac{r^j_l}{-i\omega-p^j_l} $, $j=1,3$, and $ i:=\sqrt{-1}$.

Equations (\ref{Eq:planewave1}) and (\ref{Eq:planewave2}) lead to the following equation
\begin{equation}
\mathbf{\Gamma}^{-1}\mathbf{L}\mathbf{F}\mathbf{V}_{0}=(\frac{\omega}{k})^2\mathbf{V}_{0},
\end{equation}
which is an eigenvalue problem with four eigenvalues and corresponding eigenvectors. The analytic plane wave solutions are obtained by solving this equation and choosing the eigenvalue $\left(\frac{\omega}{k}\right)^2$ and eigenvector $\mathbf{V}_{0}$.

\section{\label{numerics}Numerical results}
We now illustrate a series of numerical examples to demonstrate the performance our RKDG solver, described in Section \ref{RKDG}, for the wave propagation in orthotropic porous media. We first examine the accuracy of our code by using exact plane waves in low frequency case.  Next, we concentrate on some point source examples both in homogeneous and heterogeneous media in low frequency.  Finally the high frequency model is verified through simulating some point source examples both in single and heterogeneous media. The coefficients of the materials  are listed in Table \ref{tab_property}, the poles $p^{x,z}_j$ and residuals  $r^{x,z}_j$ used in the high frequency tests are listed in Table \ref{Tb_pol_res}. In the implementation of numerics, without loss of generality, the principal directions of the material are assumed to be coincide with the global coordinate axes.

\begin{table}[htp]
\centering
\begin{tabular}{p{20pt} p{60pt} p{50pt} p{50pt} p{50pt} p{50pt} p{60pt} }
\multicolumn{7}{p{400pt}}{}\\
\toprule
  $¡¡$¡¡& $¡¡$ & $Sandstone$  & $Glass/epoxy$ & $Sandstone$ & $Shale$ &$Glass1/epoxy1$ \\
  $ $  & $  $ & $(orthotropic)$ & $(orthotropic)$ & $(isotropic)$ & $(isotropic)$& $(orthotropic)$ \\
\midrule
  $Basic \ \ \ \ \ \ \ \ \ \  properties$¡¡&¡¡$   $  &  $    $  &  $   $ &  $¡¡$ &  $¡¡$  \\
\hline
  $K_{s}$ & $(Gpa)$  & 80 & 40 & 40 &7.6 & 40 \\
  $\rho_{s}$ & $(kg/m^3)$ & 2500 & 1815 &2500 & 2210 &1815                  \\
  $c_{11}  $ & $(Gpa)$    & 71.8 & 39.4 & 36  & 11.9 &39.4                \\
  $c_{12}  $ & $(Gpa)$    & 3.2  & 1.2  & 12  & 3.96 &1.0                    \\
  $c_{13}  $ & $(Gpa)$    & 1.2  & 1.2  & 12  & 3.96 &5.8                     \\
  $c_{33}  $ & $(Gpa)$    & 53.4 & 13.1 & 36  & 11.9 &13.1                      \\
  $c_{55}  $ & $(Gpa)$    & 26.1 & 3.0  & 12  & 3.96 &3.0                       \\
  $\phi    $ & $     $    & 0.2  & 0.2  & 0.2 & 0.16 &0.2                       \\
  $\kappa_{1}$ &$(10^{-15}m^2)$  & 600  & 600 & 600 & 100 &600                      \\
  $\kappa_{3}$ &$(10^{-15}m^2)$  & 100  & 100 & 600 & 100 &100                     \\
  $\alpha_{\infty1}$    & $    $     &  2   & 2    & 2   &  2  &2.0                           \\
  $\alpha_{\infty3}$    & $    $     &  3.6 & 3.6  & 2   &  2  &3.6                           \\
  $K_{f}$    & $(Gpa)$    & 2.5 & 2.5   & 2.5 & 2.5 &2.5                         \\
  $\rho_{f}$ & $(kg/m^3)$ &1040 & 1040  & 1040& 1040 &1040                        \\
  $\eta$ & $(10^{-3}kg/m.s)$ & 1 & 1    & 0   & 0     &1                        \\
  $\Lambda_1$ &(m) & $6.93\times 10^{-6}$ &$- $ & $ -$ &$ -$ & $6.93\times 10^{-6}$\\
  $\Lambda_3$ &(m) & $3.79\times 10^{-6}$ &$- $ & $ -$ &$ -$ & $3.79\times 10^{-6}$\\
\toprule
  $Derived  \  \ \ \ quantities$¡¡&¡¡$    $  &  $    $  &  $   $ &  $¡¡$ &  $¡¡$ \\
\midrule
  $c_{pf_{1}}$ & $(m/s)$ & 6000 & 5240 & 4250 & 2480 &5230  \\
  $c_{pf_{3}}$ & $(m/s)$ & 5260 & 3580 & 4250 & 2480 &3580 \\
  $c_{s_{1}}$  & $(m/s)$ & 3480 & 1370 & 2390 & 1430 &1360  \\
  $c_{s_{3}}$  & $(m/s)$ & 3520 & 1390 & 2390 & 1430 &1380  \\
  $c_{ps_{1}}$ & $(m/s)$ & 1030  & 975  & 1020 & 1130&900   \\
  $c_{ps_{3}}$ & $(m/s)$ & 746  & 604  & 1020 & 1130 &530  \\
  $\tau_{d_{1}}$ &  $(\mu s)$ & 5.95  & 5.58  & -- & -- &--  \\
  $\tau_{d_{3}}$ &  $(\mu s)$ & 1.82  & 1.81  & -- & -- &--  \\
\bottomrule
\end{tabular}
\caption{Material properties, taken from \cite{Yvonne}}
\label{tab_property}
\end{table}
\subsection{Plane wave convergence results for LF-Biot equations - the inviscid case}
As the first example, we verify the high order accuracy of the RKDG method by varying the mesh size. We choose the inviscid orthotropic sandstone ($\eta=0$) to be the computational material, whose properties are listed in Table \ref{tab_property}. In the inviscid case, the viscous dissipation term is omitted, and the system (\ref{eq:lowsyms}) reduces to a homogeneous model.

The computational domain is $\Omega:=[-4,~4]m\times[-4,~4]m$ in the $xz$ plane, which is partitioned with a series of uniform meshes of $100-200-400-800$. The analytic plane waves solutions with the incident wave direction $(l_{x},~l_{z}) =\left(\frac{\sqrt{3}}{2},~\frac{-1}{2}\right)$ are obtained by following Section \ref{Analytic} and the angular frequency is fixed at $10^4 rad/s$. Boundary conditions are handled by setting the ghost cells with the exact value of the analytic plane waves, the parameters of the approximation solution in ghost cells are assigned by using the $L^2$ projection of the boundary data. The initial conditions are obtained by setting $t=0$ of the analytic plane waves. The CFL number equals to $0.3$ for the first order polynomial approximation and 0.18 for the second order polynomial. The total simulation time is $T=2\pi\times10^{-4}s$, one period of the wave at this frequency. In this example, limiters are not used because the solution is smooth.

Since the stress components are about seven orders of magnitude larger than the velocities, we use 1-norm:=$\frac{\int_{\Omega}\|\mathbf{Q}_{ij,n}-\mathbf{Q}_{ij,e}\|_Edxdz}{\int_{\Omega}dxdz}$ introduced in \cite{Yvonne} to measure the errors between our numerical results and the analysis plane wave solutions.  The energy norm $\|\cdot\|_E$ has the explicit form of
\begin{equation}
\|\mathbf{Q}\|_E:=\sqrt{\mathbf{Q}^HE\mathbf{Q}},
\end{equation}
where the superscript $H$ presents the complex conjugate transpose, and the matrix $E$ is positive. For the inviscid case, the analytic plane wave solutions have the form of $\mathbf{Q }=\mathbf{Q}_{0} \exp\left(i(k_{x}x+k_{z}z-wt)\right)$ with real wave number $k$ and amplitude $\mathbf{Q}_0$, which can be separated into follows:\\
\begin{equation*}
\mathbf{Q}^r=\mathbf{Q}_0 \cos(k_{x}x+k_{z}z-wt),
\end{equation*}
\begin{equation*}
\mathbf{Q}^i=\mathbf{Q}_0 \sin(k_{x}x+k_{z}z-wt),
\end{equation*}
where $\mathbf{Q}^r$ and $\mathbf{Q}^i$ are real part and imaginary part of $\mathbf{Q}$. Clearly, for the energy norm, there exists the fact that
\begin{equation}
\mathbf{Q}^HE\mathbf{Q}=\sum_{m=1}^{8}\mathbf{E}_{m,m}\left(\mathbf{Q}_{m,m}^{r^2}
+\mathbf{Q}_{m,m}^{i^2}\right)+2\sum_{m,l=1,i\neq j}^8 \left(\mathbf{Q}_m^r \mathbf{Q}_l^r + \mathbf{Q}_m^i\mathbf{Q}_l^i\right)=(\mathbf{Q}^r)^HE\mathbf{Q}^r+(\mathbf{Q}^i)^HE\mathbf{Q}^i,
\end{equation}
so it follows that
\begin{equation}
\|\mathbf{Q}\|_E =\|\mathbf{Q}^{r}\|_E+\|\mathbf{Q}^{i}\|_E.
\end{equation}

\begin{table}
\centering
\begin{tabular}{ p{60pt}  p{60pt} p{30pt} p{20pt} p{60pt} p{30pt} }
\toprule
  & \multicolumn{2}{c}{$P^1$}& & \multicolumn{2}{c}{$P^2$} \\
\cline{2-3} \cline{5-6}
  $ Mesh$  & Error &Order &    & Error &Order \\
\midrule
   $100\times100$ &1.78E-01  & --    &  & 1.7574E-03  & --  \\
   $200\times200$ &4.41E-02  & 2.01  &  & 2.1792E-04  & 3.01  \\
   $400\times400$ &1.10E-02  & 2.00  &  & 2.7132E-05  & 3.01   \\
   $800\times800$ &2.76E-03  & 2.00  &  & 3.3856E-06  & 3.00   \\
\bottomrule
\end{tabular}
\caption{$\mathbf{\|Q\|}_{E}$ error and order of the inviscid case by DG method}
\label{Tb_inviscid}
\end{table}

In the implementation of the simulation, we calculate the real and imaginary energy errors at every cell and then add them up to obtain the total error. Table \ref{Tb_inviscid} shows that the proposed RKDG methods have accuracy order of $n + 1$ order  for $P^n$ approximations with $n$ = 1, 2, and therefore they are optimal.

\subsection{\label{sec_plane_wave}Plane wave convergence results for LF-Biot equations - the viscous case }

In this section, we consider the example that the fluid inside the porous is viscous, where the system (\ref{eq:lowsyms}) becomes a dissipative system with the viscous property being included. Because the dissipation term has its own decay timescale and is independent of the computational grid, the equation (\ref{eq:lowsyms}) is a stiff system of partial differential equation for some large scale cell sizes. Handling the viscous term is a difficult problem in the numerical simulation. In \cite{Yvonne, Puente, Lombard2}, a splitting method is used to deal with the viscous term, in which the dissipation term is exactly solved by solving an ODE system. Thanks to the stability property of the RKDG method, the computational time step is automatically smaller than the viscous decay time so that we could use the direct method to calculate the viscous term.

The computational domain is $[-0.6,~ 0.6]m\times[-0.6,~0.6]m$ in the $xz$ plane and the computational material is the viscous orthotropic sandstone with viscousity $\eta=0.001$.  In this example, the domain is partitioned with a series of uniform meshes of 100-200-400-800.  We choose the incident direction to be $(l_{x},~l_{z}) =(1,~0)$  and the angular frequency of the plane waves is fixed at $10^4 rad/s$.  The initial data are obtained by setting $t=0$ for the analytic plane waves, and the numerical solution is obtained by using the $L^2$ projection of the initial data. Boundary conditions are handled by setting the ghost cells with the exact value of the analytic plane waves, and the parameters of the approximation solution in ghost cells are assigned by using the $L^2$ projection of the boundary data. The final time is $T_1=1.25\times10^{-4}s$, which is $1.25$ periods of the fast wave and S wave. The CFL numbers equal to $0.3$ for the first order polynomial approximation and $0.18$ for the second order polynomial approximation.

Similarly, assume that the analytic plane wave solutions for the viscous case have the form of $\mathbf{Q }=\mathbf{Q}_{0} \exp(i(kl_{x}x+kl_{z}z-\omega t))$, the wave number $k$ and amplitude $\mathbf{Q}_0$ are complex numbers since the dissipation term is included.  Denote $\mathbf{Q}_0=a+ib$ and $k=c+id$.  Then we can obtain the real and imaginary as follows
\begin{equation*}
\mathbf{Q}^r=[a\cos(cxl_x+czl_z-\omega t)-b\sin(cxl_x+czl_z-\omega t)]e^{-d(xl_x+zl_z)},
\end{equation*}
\begin{equation*}
\mathbf{Q}^i=[a\sin(cxl_x+czl_z-\omega t)+b\sin(cxl_x+czl_z-\omega t)]e^{-d(xl_x+zl_z)},
\end{equation*}
where $i$ is the imaginary unit, $a, \ b, \ c, \ d$ are real numbers. We compute the error at every cell and then add them up to get the total error of the problem.

\begin{table}
\centering
\begin{tabular}{  p{60pt} p{60pt} p{30pt}  p{20pt} p{60pt} p{30pt} }
\toprule
  & \multicolumn{2}{c}{$P^1$}&  &\multicolumn{2}{c}{$P^2$}\\
\cline{2-3} \cline{5-6}
       $Mesh$  & Error &Order &  & Error &Order  \\
\midrule
    $100\times100$ & 1.49E-01  & --   &  & 8.55E-04  & --   \\
    $200\times200$ & 3.78E-02  & 1.98 &  & 1.09E-04  & 2.97 \\
    $400\times400$ & 9.51E-03  & 1.99 &  & 1.38E-05  & 2.99 \\
    $800\times800$ & 2.39E-03  & 2.00 &  & 1.75E-06  & 2.98  \\

\bottomrule

\end{tabular}
\caption{$\mathbf{\|Q\|}_{E}$ error and order of the viscous case by DG method}
\label{Tb_vis}
\end{table}

As in the inviscid example, we use the 1-norm to measure the errors between the numerical results and analytic plane waves. The results are presented in Table \ref{Tb_vis}. From the results, we can see that $n+1$  accuracy order is achieved for $P^n$ approximations of the RKDG method.

\subsection{Point source for LF-Biot equations}
In this example, we present the results of the point source located in a homogeneous orthotropic medium. The poroelastic materials are orthotropic sandstone and epoxy-glass whose properties are given in Table \ref{tab_property}. The computational domain is a square $\Omega=[-9.35,~ 9.35]m \times[-9.35,~ 9.35]m$ and is centered by a point source of Ricker wavelet RW($t$),
\begin{equation}
\label{Eq:ricker}
\text{RW}(t)=
\begin{cases}
(1-2\pi^2 f_{src}^2(t-t_d)^2)\text{exp}(-\pi^2 f_{src}^2(t-t_d)^2) &\mbox{\text{if}~ $0\leq t  \leq t_d$} ,\\
0 & \mbox{otherwise},
\end{cases}
\end{equation}
where $f_{src}$ is the peak frequency of the wavelet, $t_d$ is the time delay. The peak frequencies of the wavelet are $f_{src}=3730$ Hz  for sandstone and $f_{src}=3135$ Hz for glass, which are below the Biot characteristic frequency $f_c$ and belong to the low frequency range.
The source with a time delay of $0.0004s$ acts on $\tau_{zz}$  and $p$ simultaneously, with peak intensity $+1 Pa \cdot m^2/s$ and $-1 Pa \cdot m^2/s$, respectively. The simulation time is $0.00156s$ for sandstone and $0.0018s$ for glass, respectively. In this example, both inviscid and viscous pore fluids are considered.

\begin{figure}[htp]
\centering
\subfloat[inviscid $v_x$]{%
   \includegraphics[width=40mm]{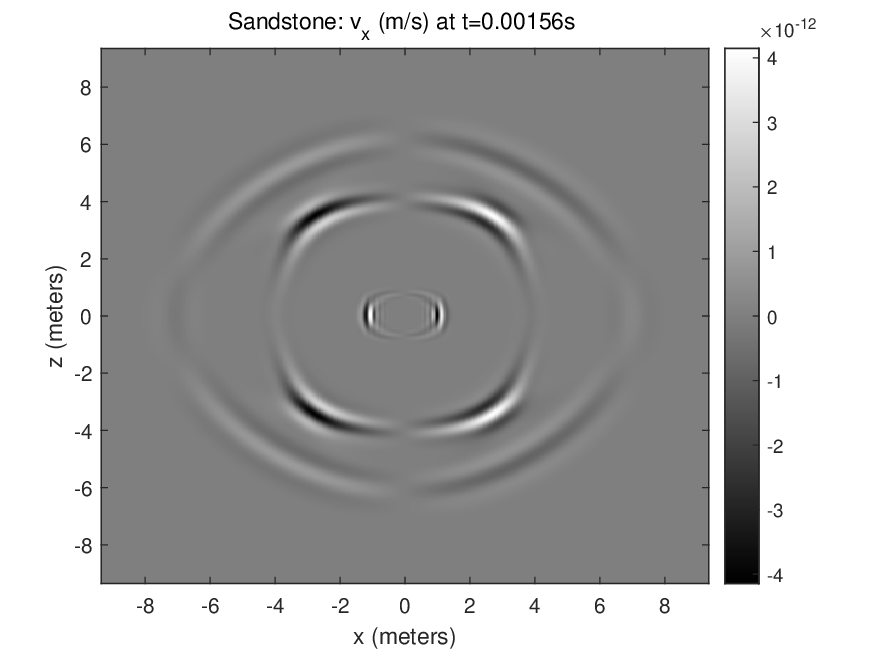}}
\subfloat[viscous $v_x$]{%
   \includegraphics[width=40mm]{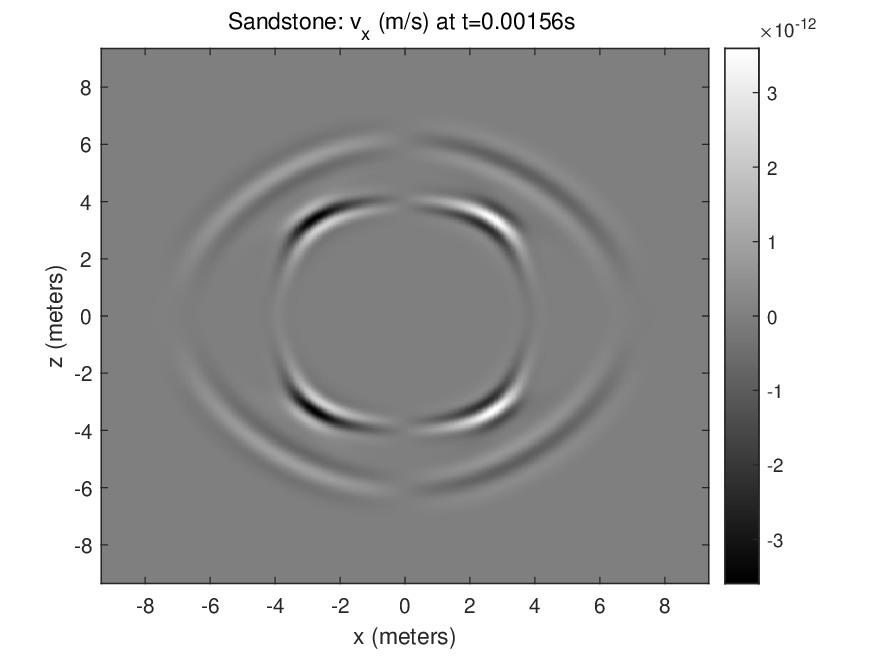}}
\subfloat[inviscid $v_z$]{%
   \includegraphics[width=40mm]{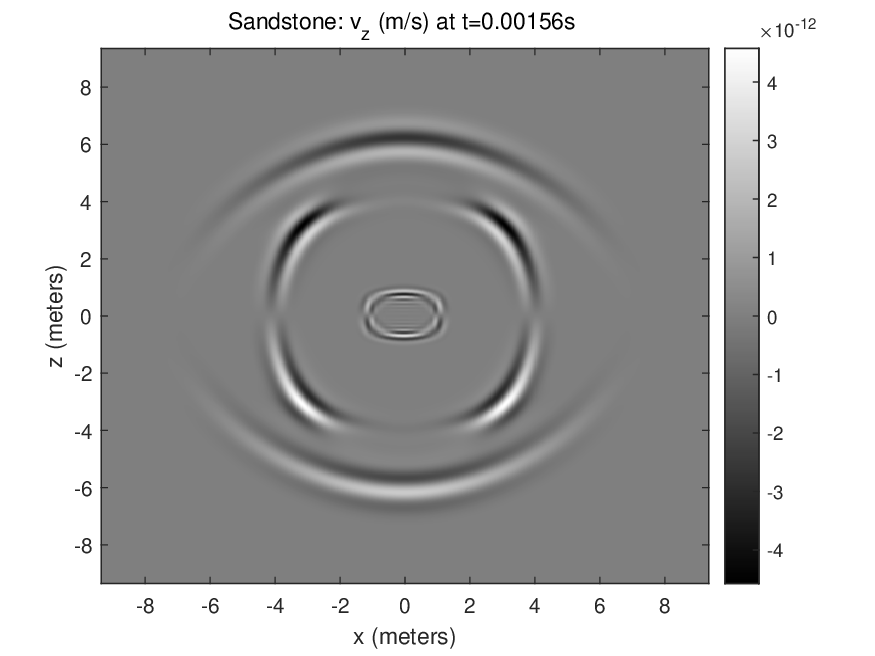}}
\subfloat[viscous $v_z$]{%
   \includegraphics[width=40mm]{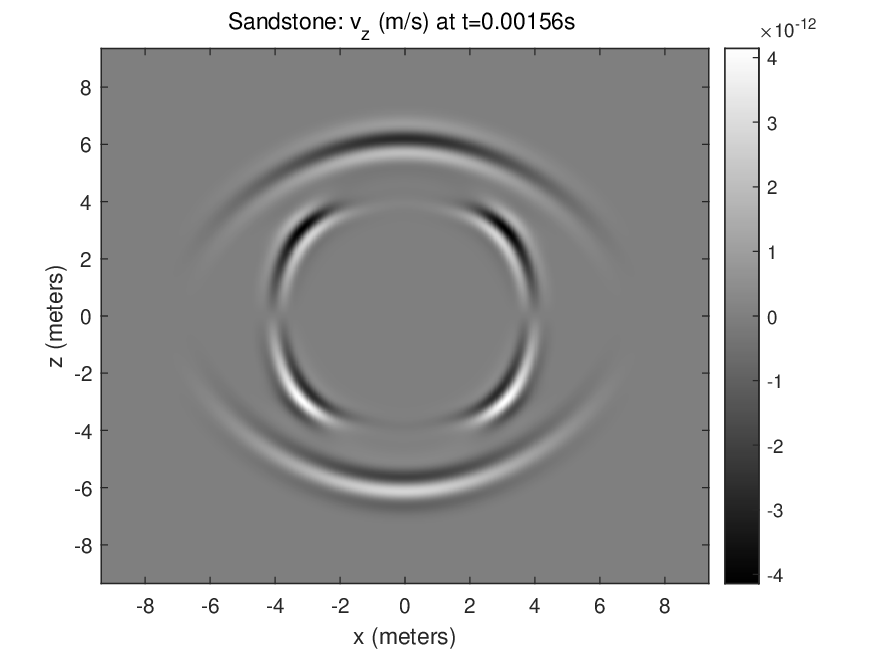}}\\
\subfloat[inviscid $v_x$]{%
   \includegraphics[width=40mm]{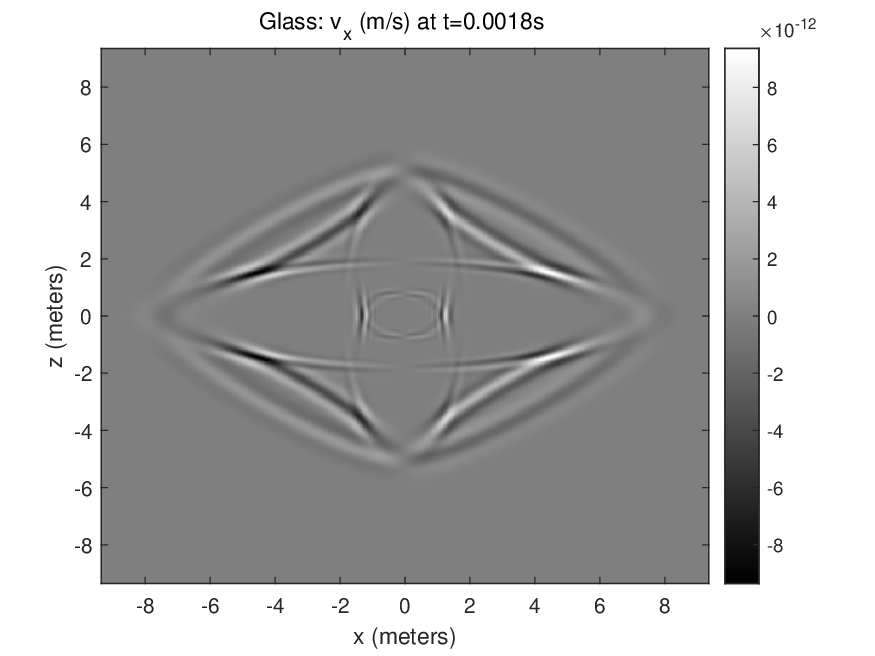}}
\subfloat[viscous $v_x$]{%
   \includegraphics[width=40mm]{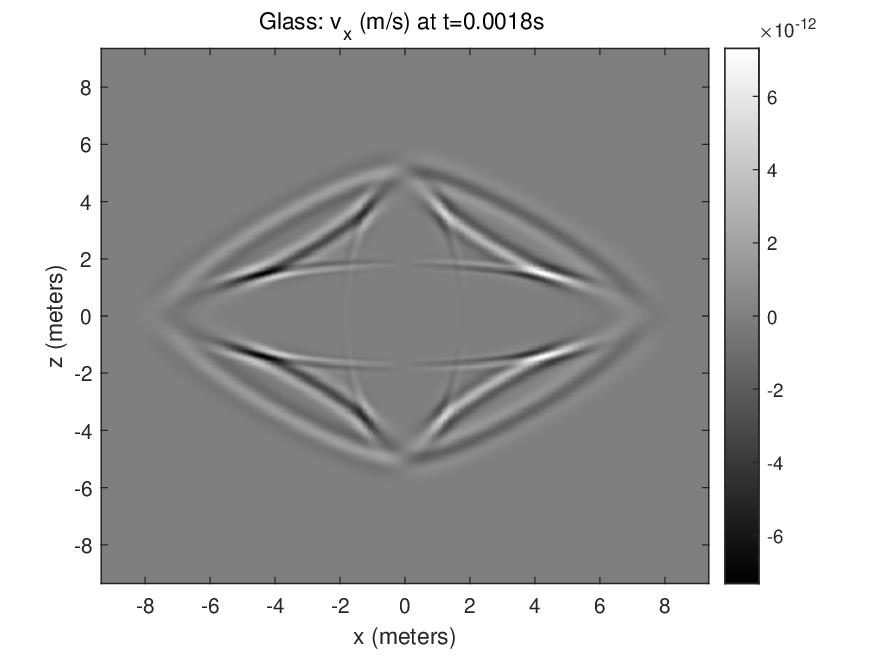}}
\subfloat[inviscid $v_z$]{%
   \includegraphics[width=40mm]{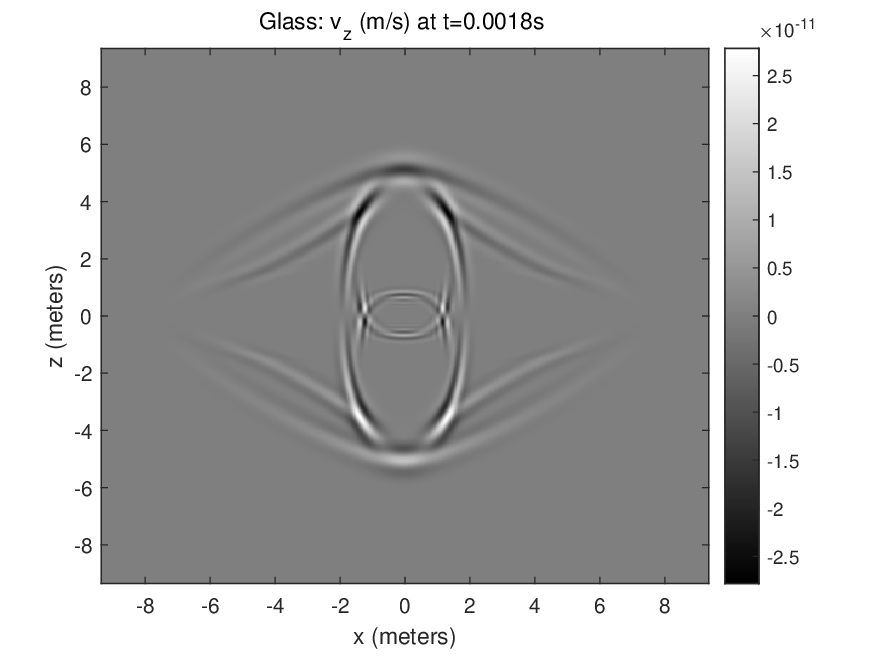}}
\subfloat[viscous $v_z$]{%
   \includegraphics[width=40mm]{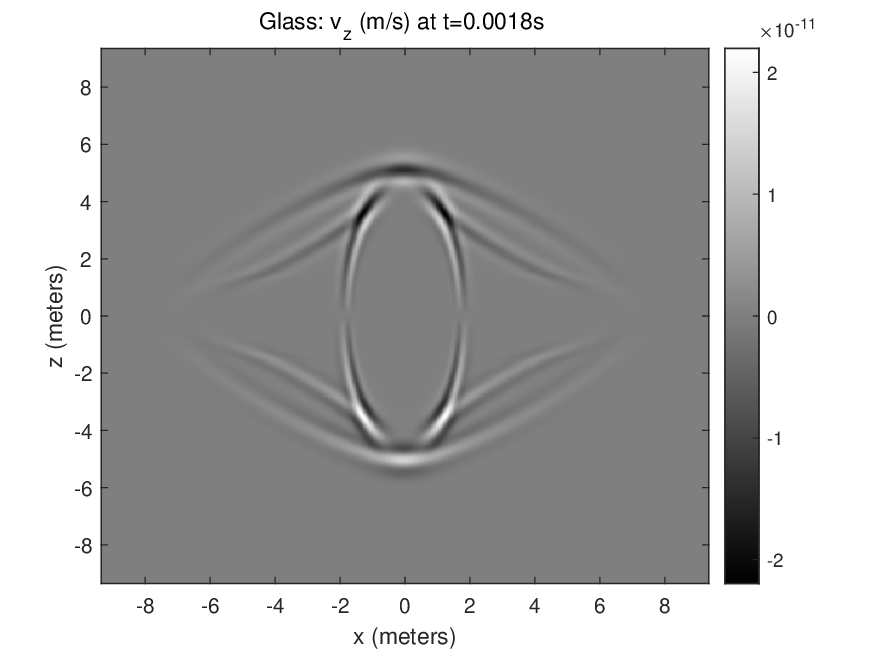}}\\
\caption{\label{fig_homog}Solid-particle velocities for point source with Ricker wavelet. (a)-(d): Sandstone with $f_{src}=3730$ Hz, (e)-(h): Glass with $f_{src}= 3135$ Hz.{\iffalse (a)(c)(e)(g) are results for inviscid case, (b)(d)(f)(h)are the viscous case.\fi} }
\end{figure}

The computation domain is covered by a uniform mesh of rectangular $501\times501$ grid cells, in which the odd number of grid cells ensures the point source to be located in a single cell. The initial condition is set to be $\mathbf{Q}=\mathbf{0}$, and the values of ghost cells are set by using outflow boundary conditions. The CFL number is $0.3$ for the first order approximation and $0.18$ for the second order approximation. In this example, the componentwise limiter with $M=50$ is used in the simulation, and the effects are the same if the local characteristic projection limiter is used. The snapshots results of the solid-particle velocities are shown in Figure \ref{fig_homog}. From the results, we can observe that the point source generate the three waves: $fast~ P~wave,~S~wave~and~slow~P~wave$. The $\it Slow\ P\ wave $ in the viscous case becomes diffusive wave centered at the source location. These results are consistent with the results of J. M. Carcione et al. \cite{Carcione}, de la Puente et al. \cite{Puente} and M. Y. Ou et al. \cite{Yvonne}.

\subsection{Heterogenous domain for LF-Biot equations}
After testing the homogeneous material with the point source, in this example we focus on the case for which the computation domain is composed of multiple materials.  The upper is isotropic shale, and the lower is isotropic sandstone whose properties are described in Table \ref{tab_property}. The waves, in this case, will reflect, transmit and interconvert into each other at the material interface, which makes the wave propagation in this multiple materials becomes more complicated than the single case.

\begin{figure}[htb]
\centering
\subfloat[$v_z$ at t=0.25]{%
   \includegraphics[width=2.5in,height=2.0in]{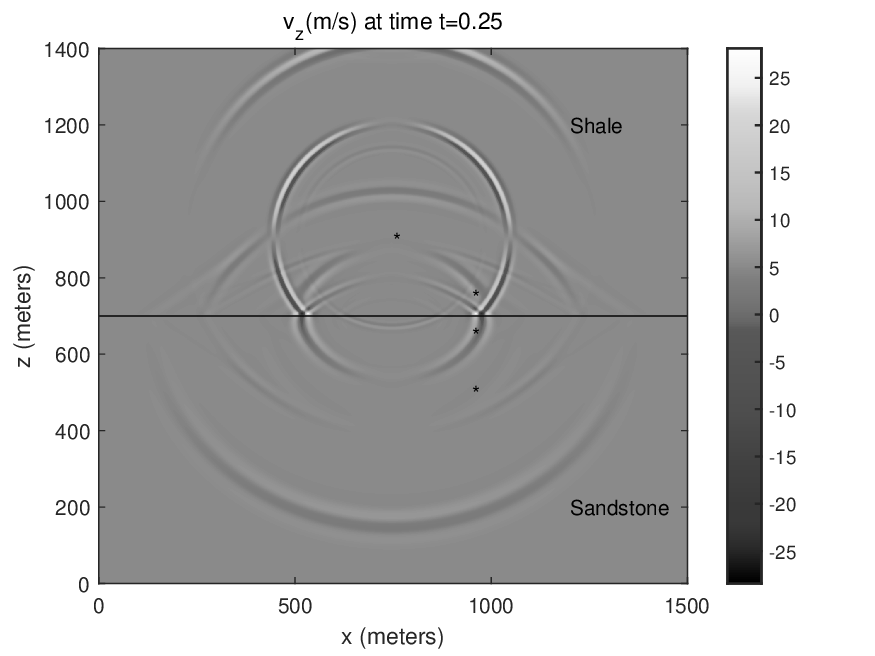}}
\subfloat[$v_z$ at t=0.25]{%
   \includegraphics[width=2.5in,height=2.0in]{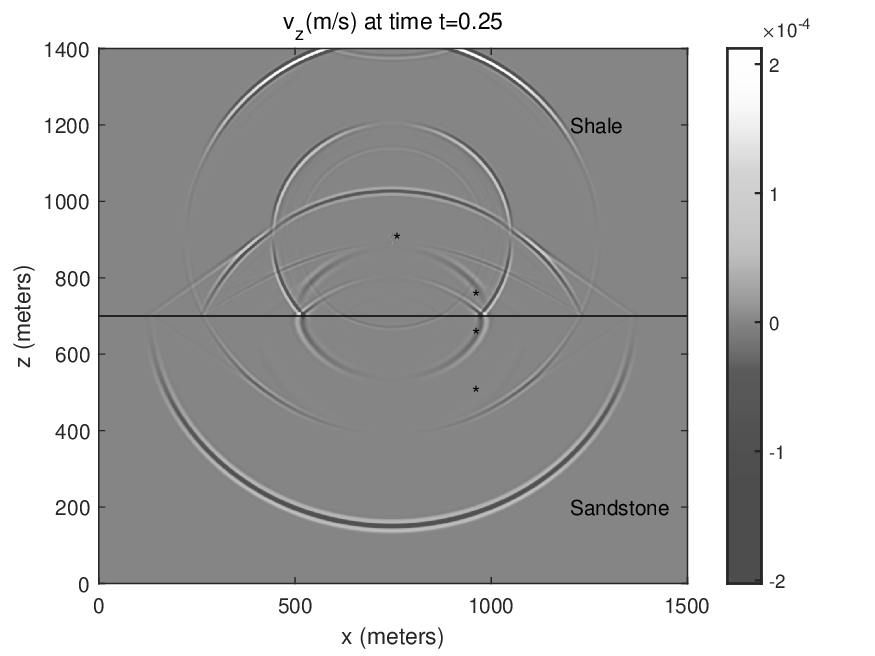}}

\caption{\label{fig_hetero} Heterogeneous domain results of $v_z$ at $t$=0.25s for point source with Ricker wavelet. (a) snapshot of $v_z$ at $t$=0.25s  for $f_{src}$=50 Hz, (b) snapshot of $v_z$ at $t$=0.25 s for $f_{src}$=4000 Hz.}
\end{figure}

The computational domain is $\Omega =[0,~1500]m\times[0,~1400]m$ in $xz$ plane with the material interface at $z=700m$. Again the point source of the Ricker wavelet is located at $(x,z)=(750m,900m)$, acting on the components of the $z$ direction normal stress $\tau_{zz}$ and the fluid pressure $p$ with peak intensities $+2.3\times10^{13} Pa\cdot m^2/s$ and $-2.3\times10^{13} Pa\cdot m^2/s$, respectively. For comparison, we choose the peak frequencies to be 50 Hz and 4000 Hz with the time delay $0.04s$ and $0.00025s$ respectively.  This choice of the source is based on de la Puente et al. \cite{Puente} and M. Y. Ou et al. \cite{Yvonne}.  In order to investigate the behavior of the solid particle velocities, three receivers of $(x_1,z_1)=(950m,750m)$, $(x_2,z_2)=(950m,650m)$ and $(x_3,z_3)=(950m,500m)$ marked with $\star$ in Figure \ref{fig_hetero} are chosen to record the solution information.

The computational domain is partitioned with $1421\times 1360$ uniform rectangle elements, where the odd number is to ensure the source is located in the interior of an element. The simulation starts with the initial condition of $\mathbf{Q}=\mathbf{0}$. Outflow boundary condition is used and the element boundaries are chosen to coincide with the interface boundary. The CFL number is $0.3$ for the first order approximation and $0.18$ for the second order approximation in this example. The total simulation time is $0.5s$ both for the two frequency. The componentwise limiter with $\text{M}=50$ is used in this example and there is no visible difference in the results when the limiter is implemented in the local characteristic fields.

\begin{figure}[htp]
\centering
\subfloat[Rec1: $v_x$ for 50 Hz]{%
   \includegraphics[width=2in,height=1.5in]{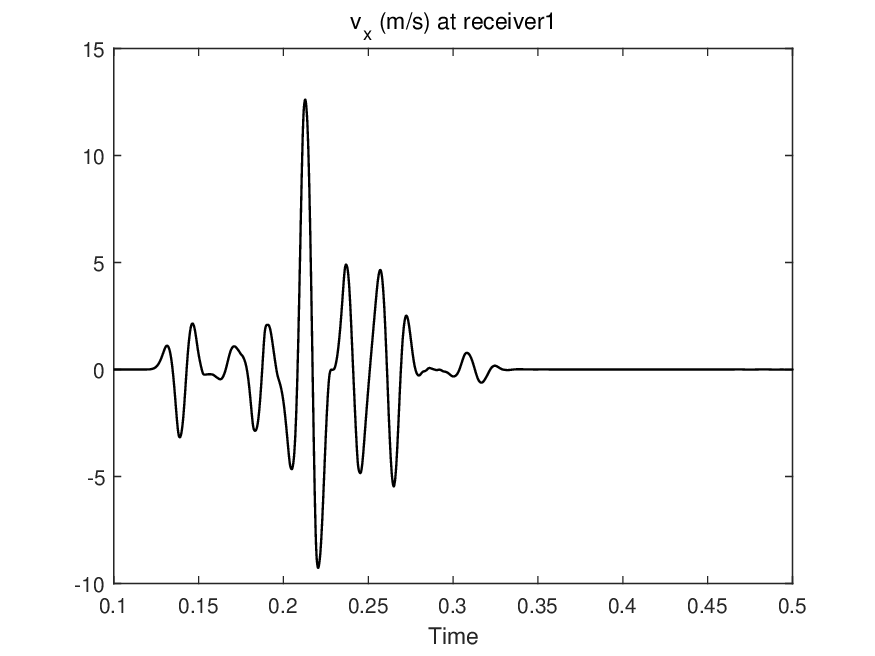}}
\subfloat[Rec2: $v_x$ for 50 Hz]{%
   \includegraphics[width=2in,height=1.5in]{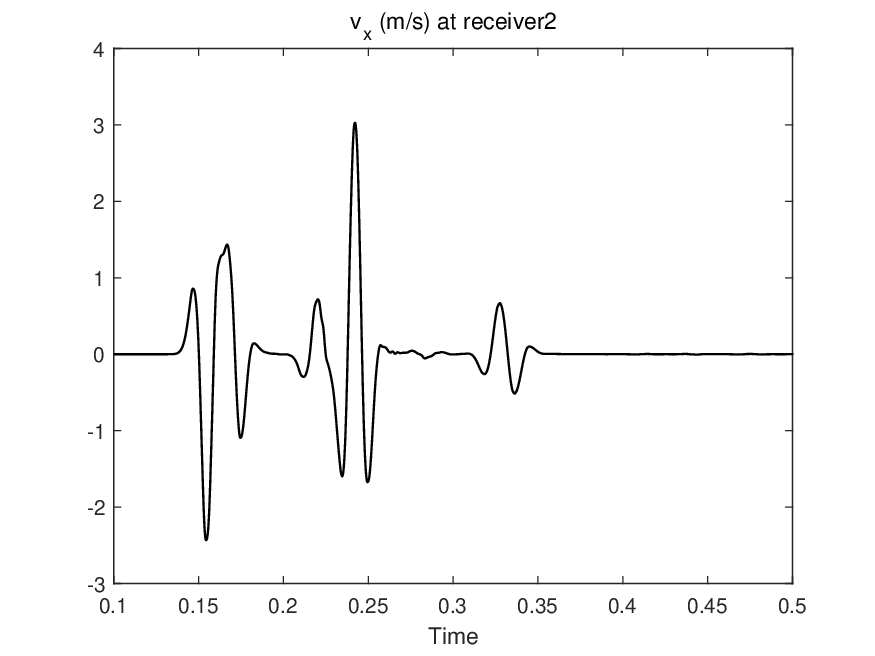}}
\subfloat[Rec3: $v_x$ for 50 Hz]{%
   \includegraphics[width=2in,height=1.5in]{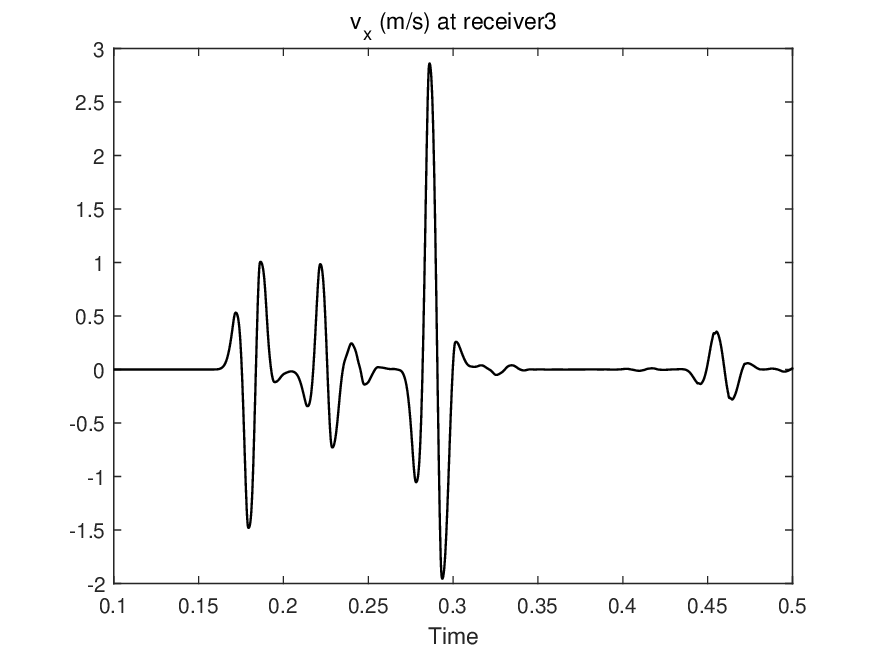}}\\
\subfloat[Rec1: $v_x$ for 4000 Hz]{%
   \includegraphics[width=2in,height=1.5in]{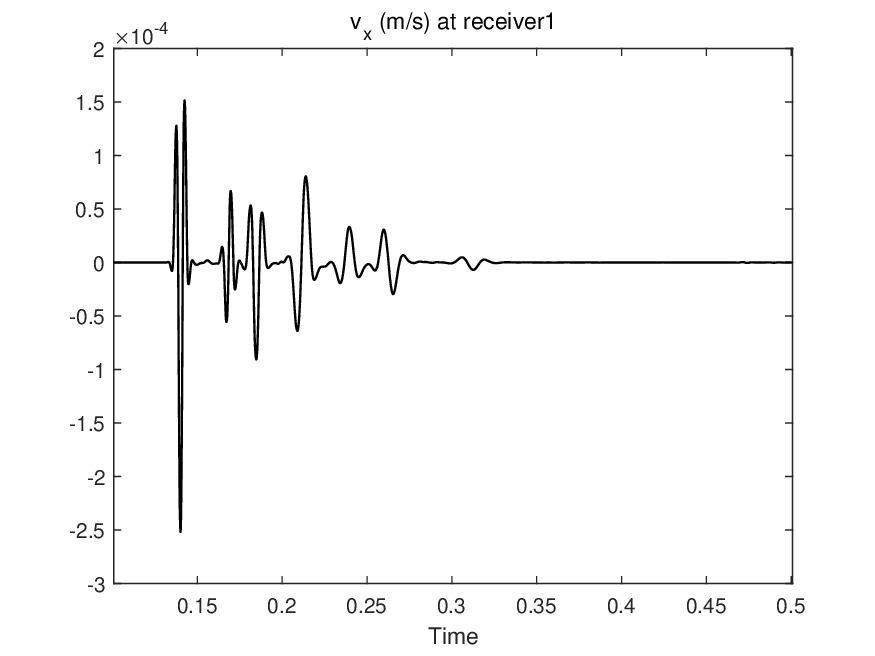}}
\subfloat[Rec2: $v_x$ for 4000 Hz]{%
   \includegraphics[width=2in,height=1.5in]{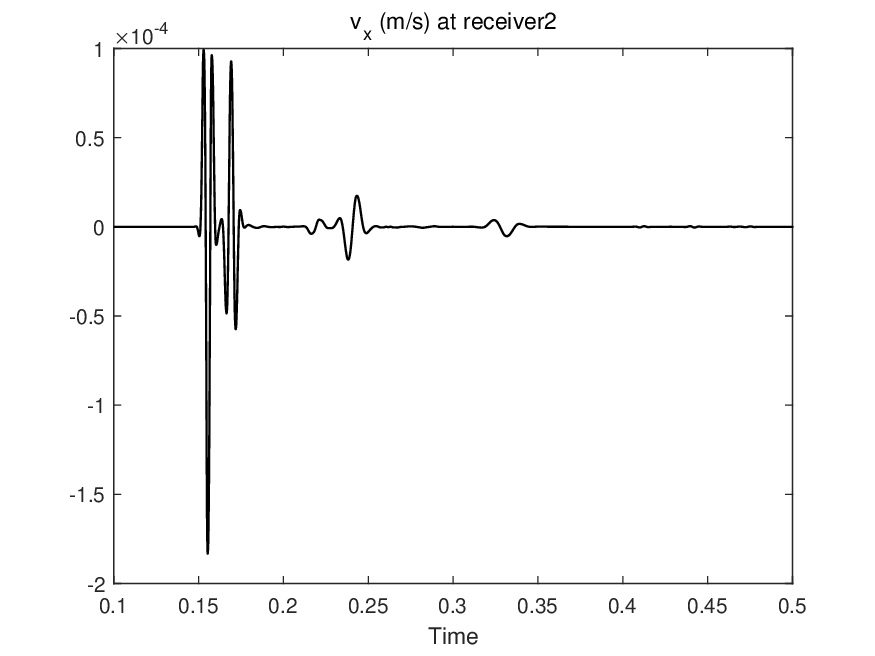}}
\subfloat[Rec3: $v_x$ for 4000 Hz]{%
   \includegraphics[width=2in,height=1.5in]{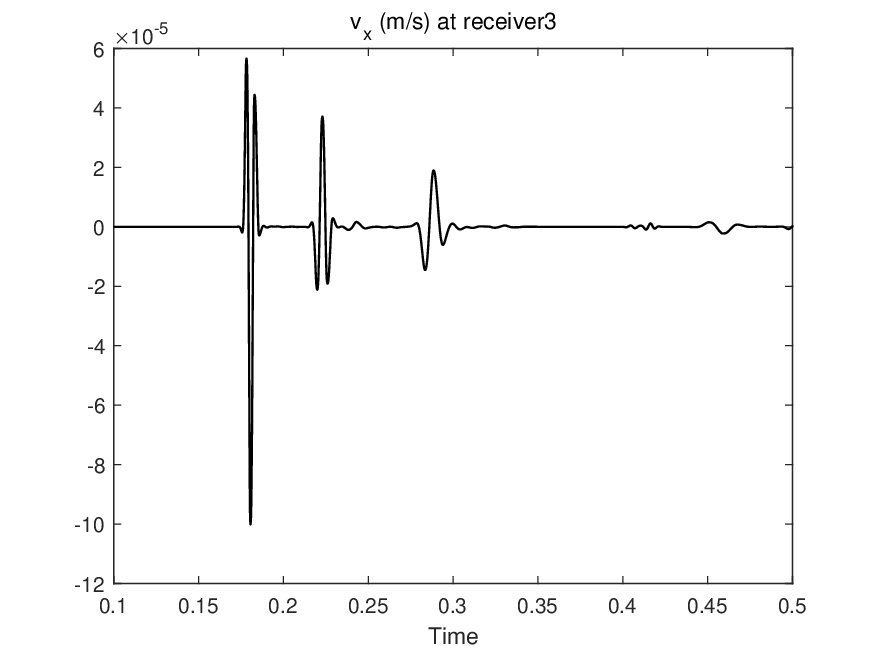}}\\
\subfloat[Rec1: $v_z$ for 50 Hz]{%
   \includegraphics[width=2in,height=1.5in]{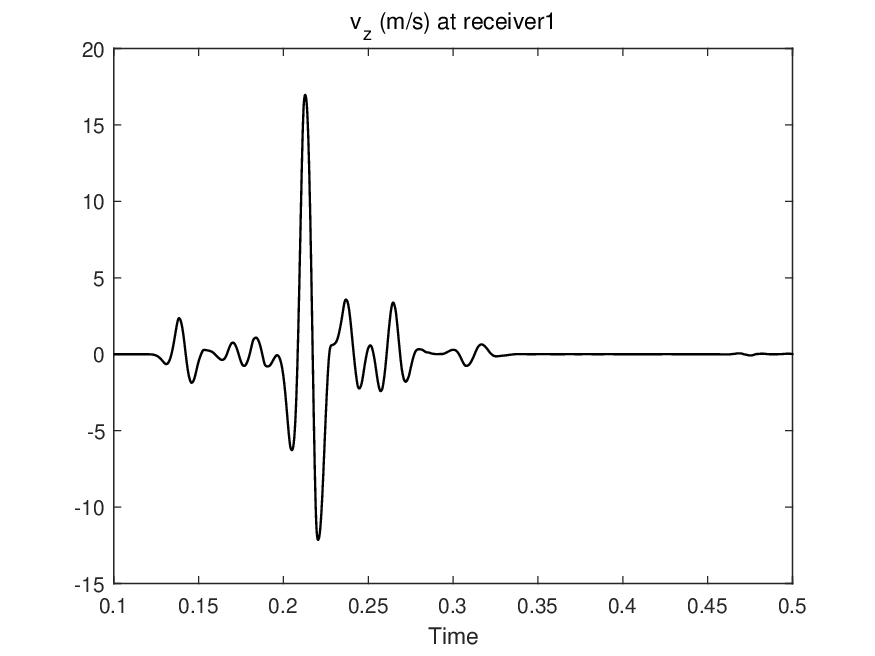}}
\subfloat[Rec2: $v_z$ for 50 Hz]{%
   \includegraphics[width=2in,height=1.5in]{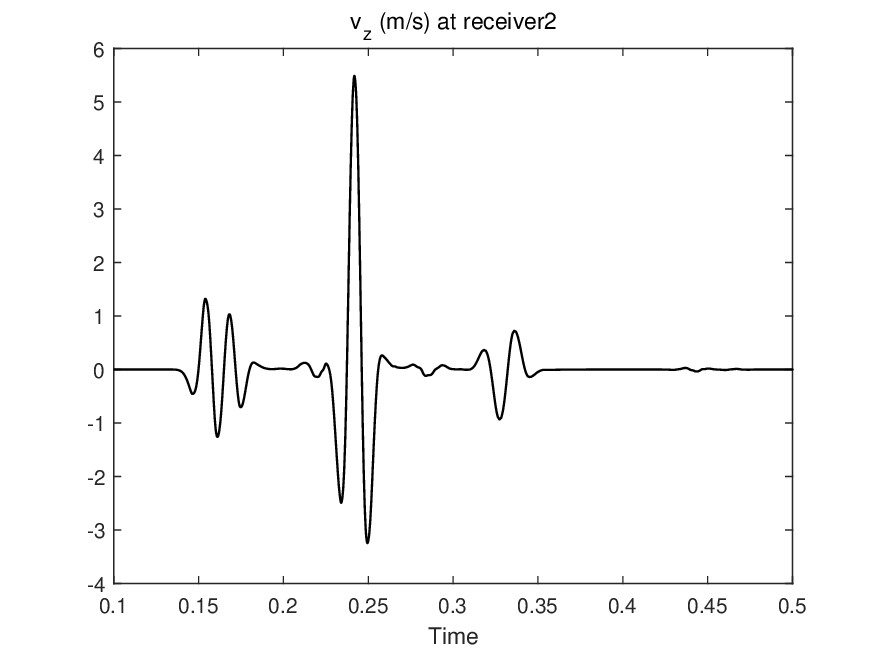}}
\subfloat[Rec3: $v_z$ for 50 Hz]{%
   \includegraphics[width=2in,height=1.5in]{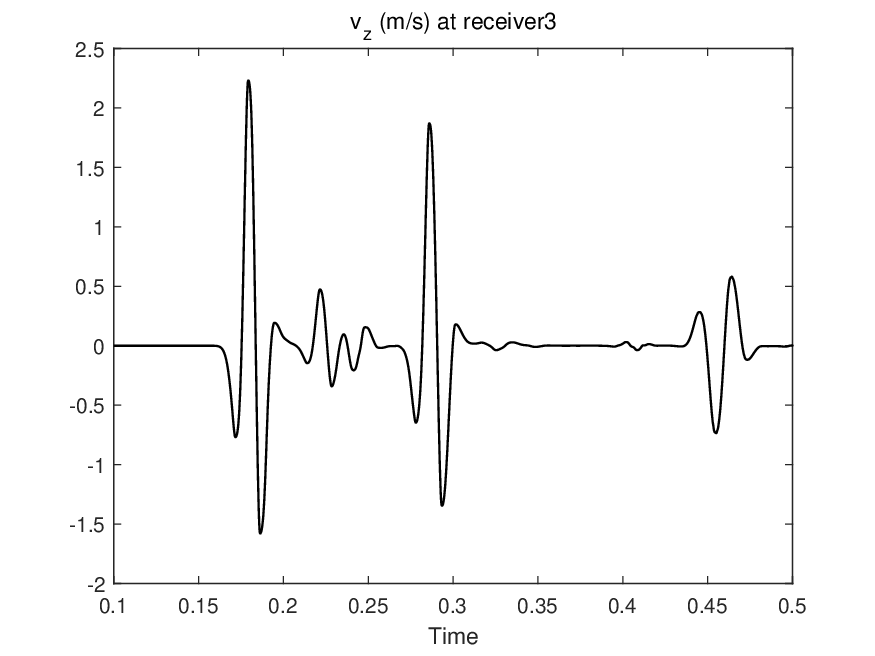}}\\
\subfloat[Rec1: $v_z$ for 4000 Hz]{%
   \includegraphics[width=2in,height=1.5in]{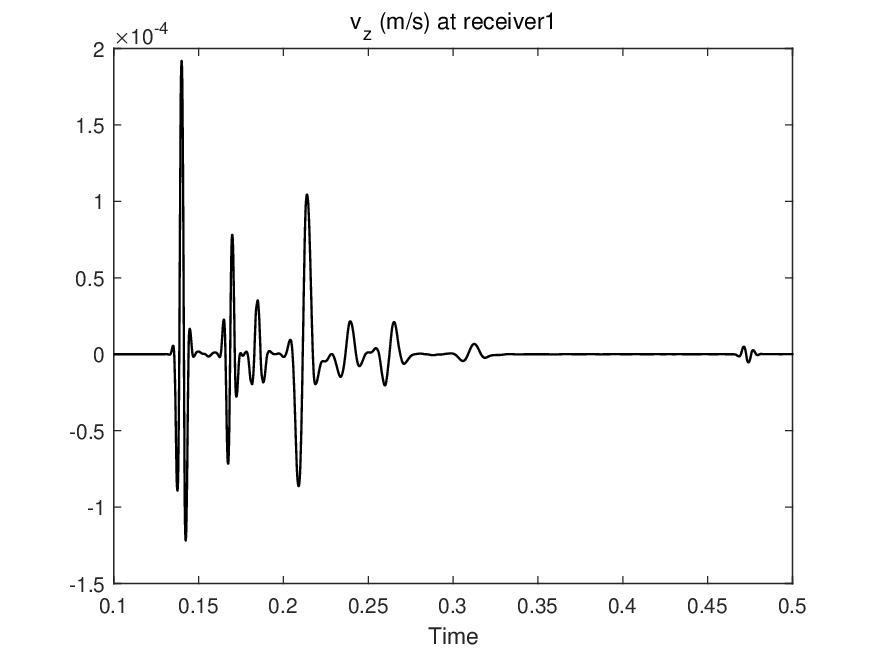}}
\subfloat[Rec2: $v_z$ for 4000 Hz]{%
   \includegraphics[width=2in,height=1.5in]{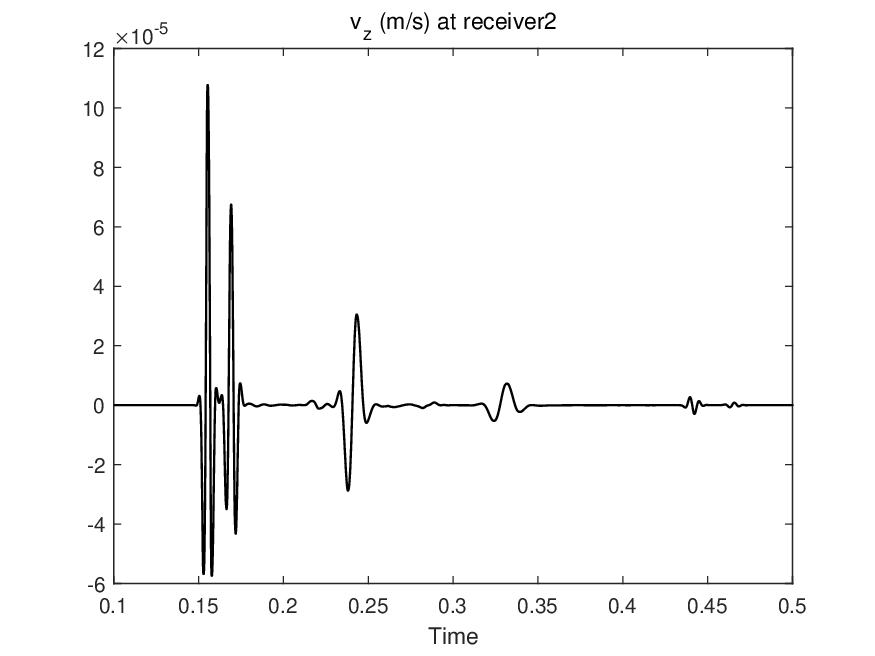}}
\subfloat[Rec3: $v_z$ for 4000 Hz]{%
   \includegraphics[width=2in,height=1.5in]{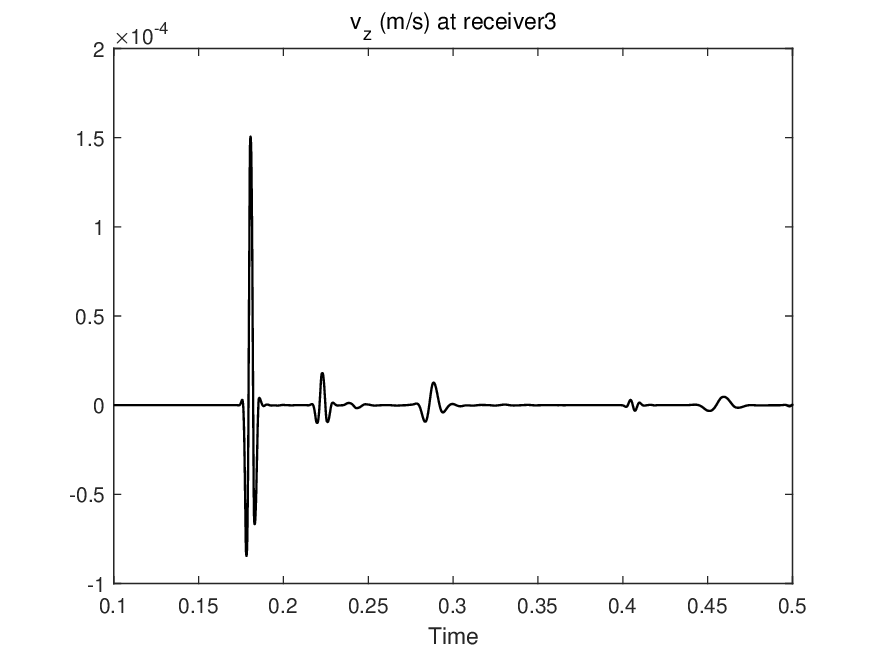}}\\
\caption{\label{fig_heteroTime}Time evolution of solid velocity at 50 Hz and 4K Hz. (a)-(c) x-direction velocities at 50 Hz, (d)-(f) x-direction velocities at 4000 Hz, (g)-(i) z-direction velocities at 50 Hz, (j)-(l) z-direction velocities at 4K Hz.}
 \label{fig:4}
\end{figure}

Figure \ref{fig_hetero} shows the snapshot results of the solid particle velocities component $v_z$ at $t=0.25s$ for both $f_{src}$=50 Hz and $f_{src}$=4000 Hz, and the relative time histories of the solid $x$ and $z$ velocities at the three receivers are shown in Figure \ref{fig_heteroTime}. For the 50 Hz case, the results are in agreement with those presented in de la Puente et al. \cite{Puente} and M. Y. Ou et al. \cite{Yvonne}.

\subsection{\label{sec_single_JKD}Single material test for Biot-JKD equations}
In this section, we run the Augmented Biot-JKD equations and the Biot Diffusive Approximation (Biot DA) equations to compare the solutions of the Biot-JKD system obtained by two different methods. The computational material is orthotropic glass1/epoxy1 with the properties listed in Table \ref{tab_property} and the critical frequencies are $f_{cx}=25.5$K Hz and $f_{cz}=85$K Hz.

In order to confirm the validation of the augmented system (\ref{aug_b}) - (\ref{aug_e}),  we choose the center frequency of the wavelet to be $f_{src}$=200K Hz to run the test listed in \cite{Lombard2}. Furthermore, to demonstrate the difference of solutions between the low frequency regime and the Augmented Biot-JKD equations in high frequency regime, we choose the center frequency of $f_{src}$=4K Hz to run another test. The poles $p^{x_j}_k$ and residues $r^{x_j}_k$  for the spectral content of each Ricker wavelet are given in Table \ref{Tb_pol_res}. The computational domain is $\Omega=[-0.15,~0.15]^2 m$ for 200K Hz and $\Omega'=[-4,~4]^2 m$ for 4K Hz, which include about 10 and 6 fast P waves respectively. The receiver at $(0.004m,0.004m)$ is chosen to record the solutions' behavior for both High and Low frequency regimes. The two domains are partitioned with $700 \times 700$ and $900 \times 900$ uniform rectangle elements respectively. The point source located at the center of the domain is used to emit cylindrical waves and is only applied on the component of $\tau_{xz}$. The point source has the form of $\text{S=g(t)h(x,z)}$, here $\text{g(t)}$ is a Ricker wavelet with the time delay $\frac{1}{f_{src}}$ and equals minus $\text{RW}$ given in (\ref{Eq:ricker}) and $\text{h(x,z)}$ is a truncated Gaussian function, via
\begin{equation}
\label{Eq:TrunGauss}
\text{h(x,z)}=
\begin{cases}
\frac{1}{\pi \Sigma^2}\text{exp}\left(-\frac{x^2+z^2}{\Sigma^2}\right) &\mbox{if \ $0\leq x^2+z^2 \leq R_0^2$},\\
0 & \mbox{otherwise},
\end{cases}
\end{equation}
with radius $R_0=6.56 \times 10^{-3}m$ and $\Sigma=3.28 \times 10^{-3}m$. The initial condition is set to be $\mathbf{Q}=\mathbf{0}$. Because the waves do not reach the edges of the computation domain, we use the zero boundary condition in the implementation of simulation.  The $\text{CFL} =0.15$ and the componentwise limiter with $\text{M}=50$ are used in this test. The total simulation time is $2.72\times10^{-5}s$  for 200K Hz and $6.73\times10^{-4}s$ for 4K Hz, which include about 5 and 3 periods of the fast P wave respectively.

\begin{figure}[htb]
\centering
\subfloat[]{%
   \includegraphics[width=2.5in,height=2.0in]{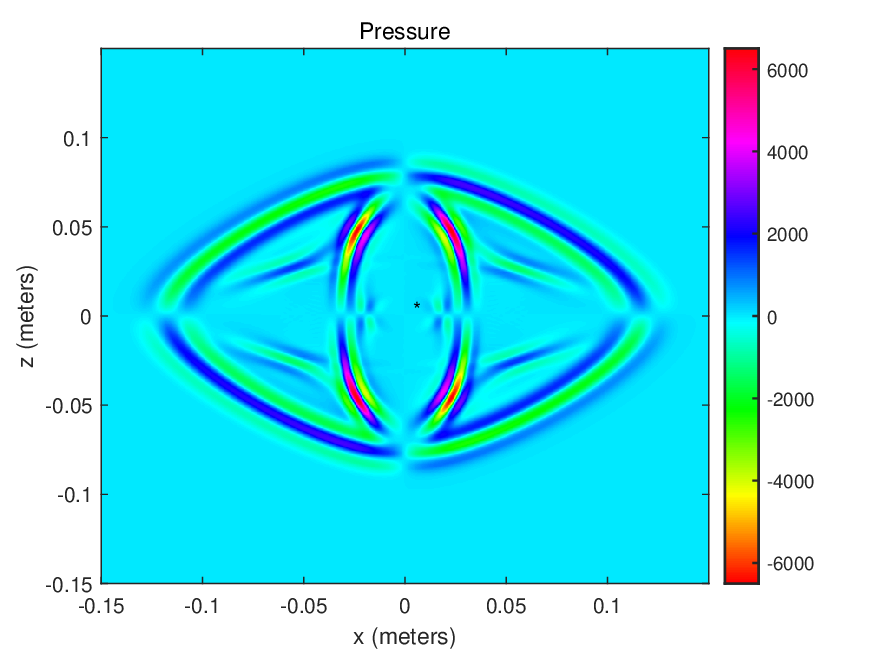}}
\subfloat[]{%
   \includegraphics[width=2.5in,height=2.0in]{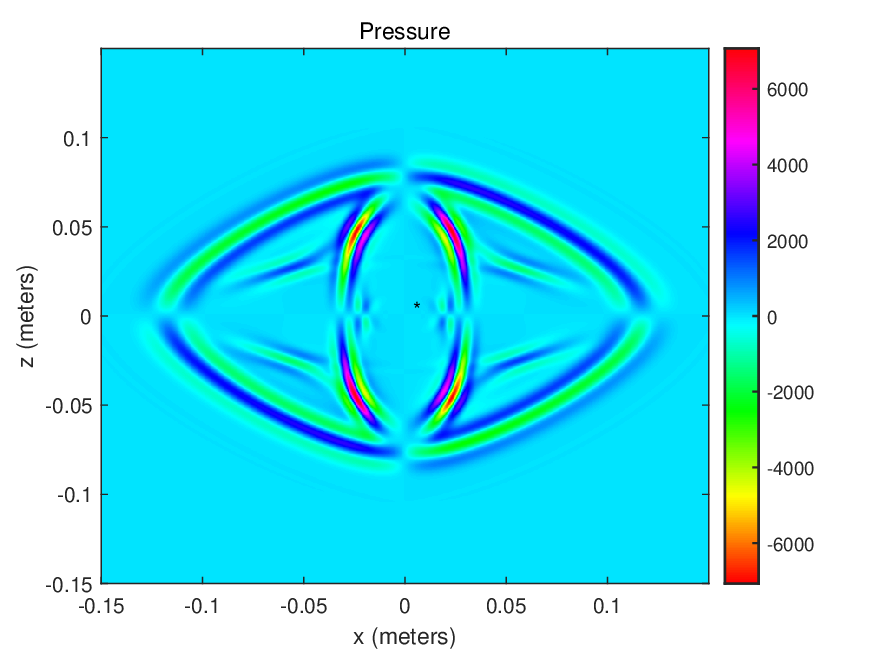}}\\
\subfloat[]{%
   \includegraphics[width=2.5in,height=2.0in]{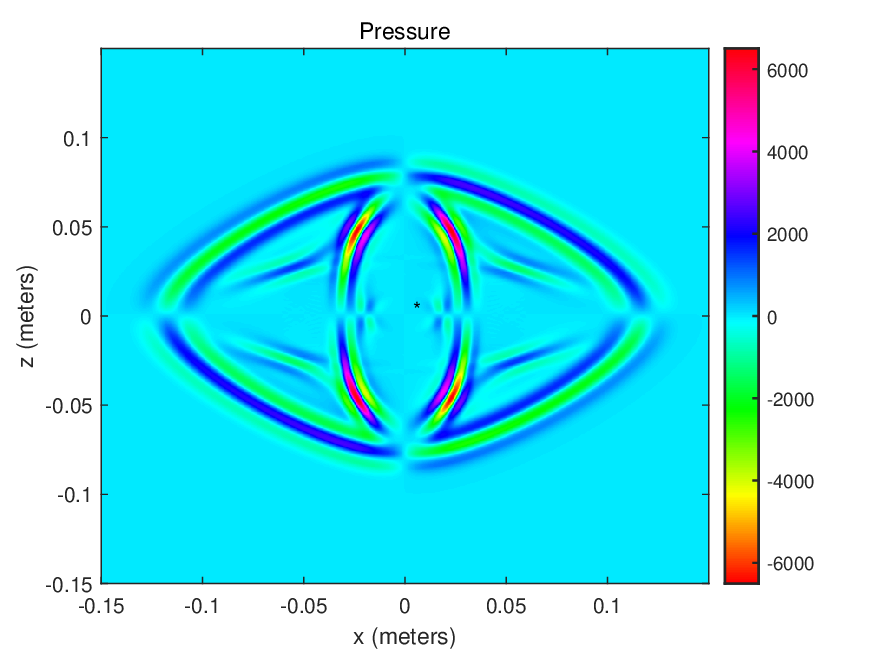}}
\subfloat[]{%
   \includegraphics[width=2.5in,height=2.0in]{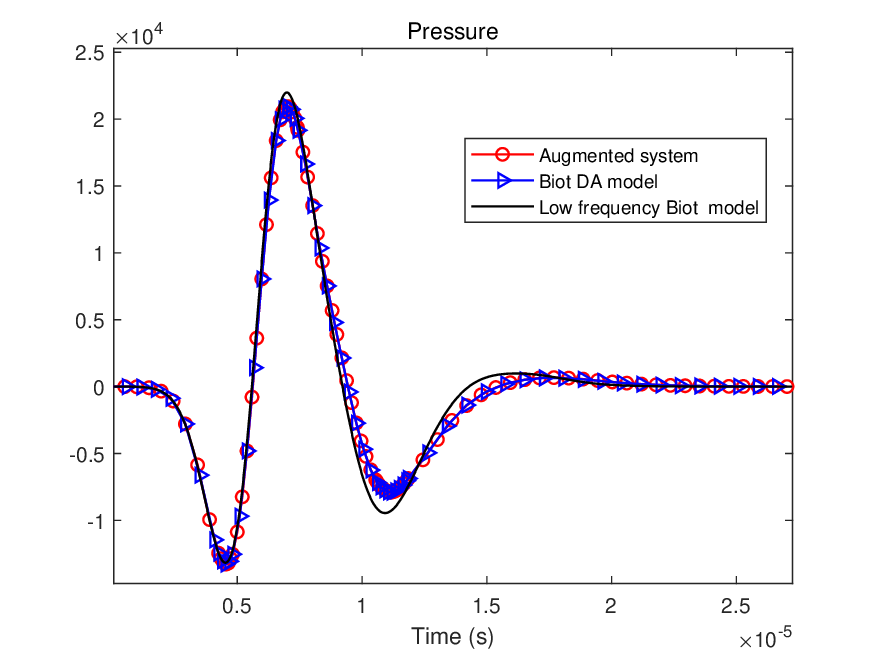}}

\caption{Pressure field of a single material of orthotropic sandstone with frequency $f_{src}$=200K Hz for LF-Biot and Biot-JKD equations with different methods. (a) Snapshot of pressure at $t= 2.72\times10^{-5}s$ using the Augmented Biot-JKD equations, (b) snapshot of pressure at $t= 2.72\times10^{-5}s$ using the LF-Biot equations, (c) snapshot of pressure at $t= 2.72\times10^{-5}s$ using the Biot-DA equations, (d) comparison of time evolution at the receiver $R$.
}
\label{fig:JKD_single_200K}
\end{figure}

Figure \ref{fig:JKD_single_200K} shows the pressure $p$ with $f_{src}$= 200K Hz for the Augmented Biot-JKD system (\ref{eq:higheqn}) and for the LF-Biot equations (\ref{eq:lowsyms}). In this test case, the solutions of Biot-JKD equations by the two different methods are indistinguishable and the one by Biot-DA is analogous to the results shown in \cite{Lombard2}.

\begin{figure}[htb]
\centering
\subfloat[]{%
   \includegraphics[width=2in,height=1.5in]{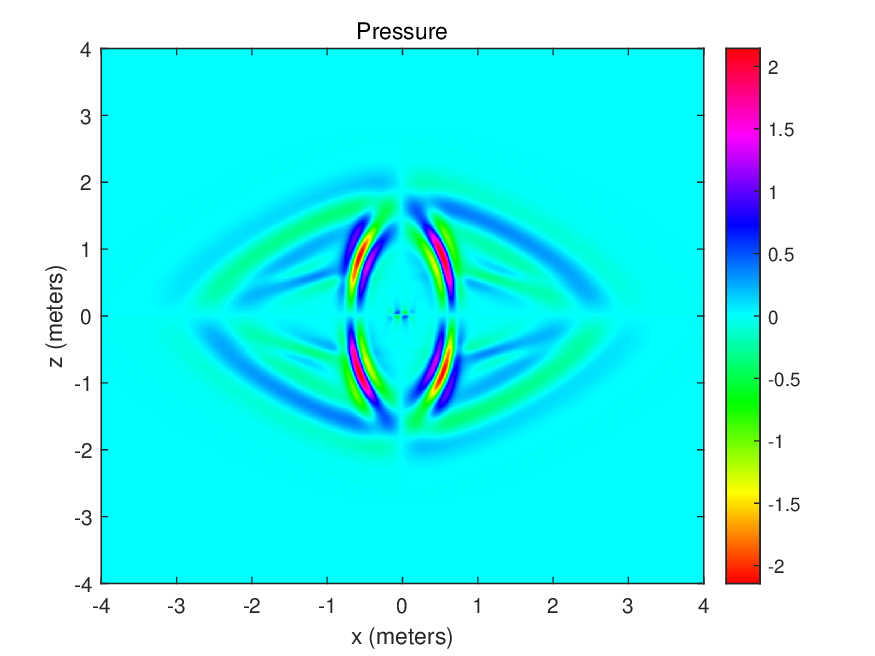}}
\subfloat[]{%
   \includegraphics[width=2in,height=1.5in]{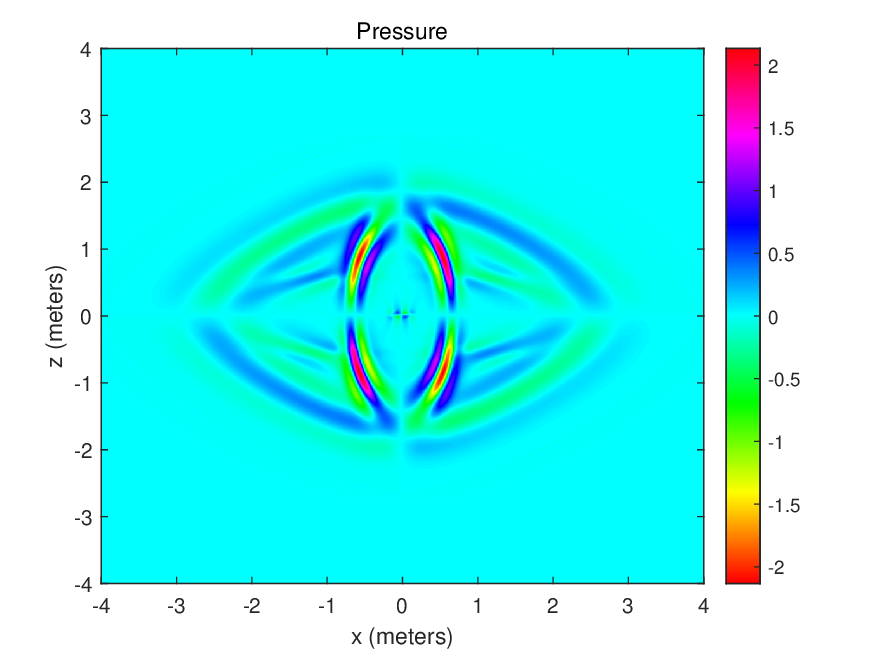}}
\subfloat[]{%
   \includegraphics[width=2in,height=1.5in]{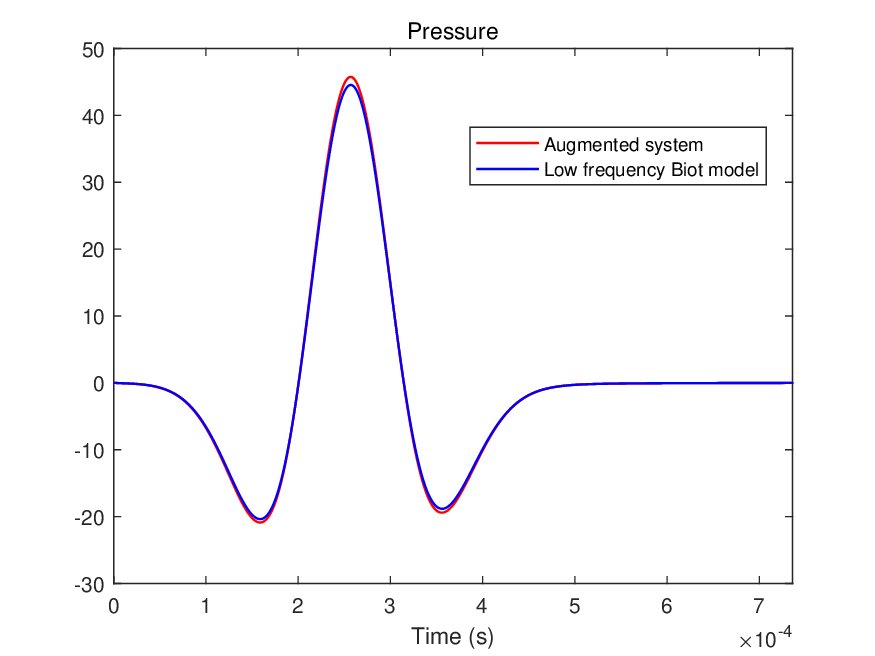}}

\caption{Pressure field of a single material of orthotropic sandstone with frequency $f_{src}$=4K Hz for LF-Biot and Biot-JKD equations. (a) Snapshot of pressure at $t= 6.73\times10^{-4}s$ using the Augmented Biot-JKD equations, (b) snapshot of pressure at $t= 6.73\times10^{-4}s$ using the LF-Biot equations, (c) comparison of time evolution at the receiver $R$.}
\label{fig:JKD_single_4K}
\end{figure}

The counterparts for $f_{src}$=4K Hz are presented in Figure \ref{fig:JKD_single_4K}, from which we can see that the LF-Biot equations are a good approximation to the Biot-JKD equations in this frequency range, which is well below the critical frequencies.

\begin{table}
\centering
\begin{tabular}{@{}ccccccc@{}}
\toprule
\multicolumn{7}{c}{200K(Hz)}\\
\hline
  &\multicolumn{2}{c}{Directions}& & &\multicolumn{2}{c}{Directions}\\
\cline{1-3} \cline{6-7}
  & $x$ ($\times 10^7$) & $z$ ($\times 10^8$) & & &$x$ ($\times 10^6$) & $z $ ($\times 10^7$)\\
\midrule
  $p_1$ & -0.035567310538715 & -0.011221462302843 & &$r_1$ & 0.006781868648686 & 0.001443060382303   \\
  $p_2$ & -0.047020191541529 & -0.013046904577717 & &$r_2$ & 0.024059080106337 & 0.005723980195443   \\
  $p_3$ & -0.069476732959186 & -0.016614322637618 & &$r_3$ & 0.047099528834055 & 0.013062384697468   \\
  $p_4$ & -0.109567304793412 & -0.023128107401683 & &$r_4$ & 0.076616812238772 & 0.025090240726876   \\
  $p_5$ & -0.183953664814239 & -0.035856379053459 & &$r_5$ & 0.124799920041847 & 0.047797031398026  \\
  $p_6$ & -0.348596359172731 & -0.065998935931129 & &$r_6$ & 0.238596041566096 & 0.103105325232612   \\
  $p_7$ & -0.905181246497858 & -0.173523259606006 & &$r_7$ & 0.671687444458139 & 0.307915072027153   \\
  $p_8$ & -7.692566436791471 & -1.510676632570611 & &$r_8$ & 6.253620575287360 & 2.892028807890477   \\
\toprule
\multicolumn{7}{c}{4K(Hz)}\\
\hline
  &\multicolumn{2}{c}{Directions}& & &\multicolumn{2}{c}{Directions}\\
\cline{1-3} \cline{6-7}
  & $x$ ($\times 10^7$) & $z$ ($\times 10^8$) & & &$x$ ($\times 10^6$) & $z $ ($\times 10^7$)\\
\midrule
  $p_1$ & -0.396110238498902 & -0.162899433433064 & &$r_1$ & 0.021601015870995 & 0.406554873958016   \\
  $p_2$ & -0.829823795820093 & -1.116705082387160 & &$r_2$ & 0.144865332656868 & 7.288457233057888   \\
  $p_3$ & -6.513380095205129 & --                 & &$r_3$ & 1.762744154711445 &  --  \\
\bottomrule
\end{tabular}
\caption{Poles and residues for material Glass1 at 200k Hz and 4K Hz}
\label{Tb_pol_res}
\end{table}

\subsection{\label{sec_couple_JKD}Couple test for Biot-JKD equations}
Based on the results of the single material for the Augmented Biot-JKD model, in this subsection, we move to consider the test that the domain consists of two isotropic materials. 

The computational domain is $\Omega =[-0.15,~0.15]^2m$ in $xz$ plane, which is separated by the interface line of $z =0 m$ and is partitioned with $700 \times 700$ uniform rectangle elements. The upper and lower domains are both isotropic materials coming from $x$ and $z$ directions of orthotropic glass1/epoxy1, where the critical frequencies are $f_{cx}=25.5$K Hz and $f_{cz}=85$K Hz, respectively. The point source $\text{S=g(t)h(x,z)}$ located at $(0m,0m)$ is used to emit cylindrical waves, where $-g(t)$ is Ricker wavelet given in (\ref{Eq:ricker}) and $h(x,z)$ is the truncated Gaussian function given in (\ref{Eq:TrunGauss}). The center frequency for the point source is 200K Hz and the poles $p^{x_{1,3}}_k$ and residues $r^{x_{1,3}}_k$ at this frequency are given in Table \ref{Tb_pol_res}. The time delay of the Ricker wavelet is $t_d=\frac{1}{f_{src}}$ and the source is only applied on the component of $\tau_{xz}$. The total simulation time is $2.72\times10^{-5}s$, which is about 5 periods long for the fast P wave. The initial condition is set to be zero and the boundary condition is chosen to be outflow boundary condition. In the implement of simulation, the $\text{CFL}= 0.15$ and componentwise limiter are used in this test.

\begin{figure}[htb]
\centering
\subfloat[]{%
   \includegraphics[width=2.5in,height=2.0in]{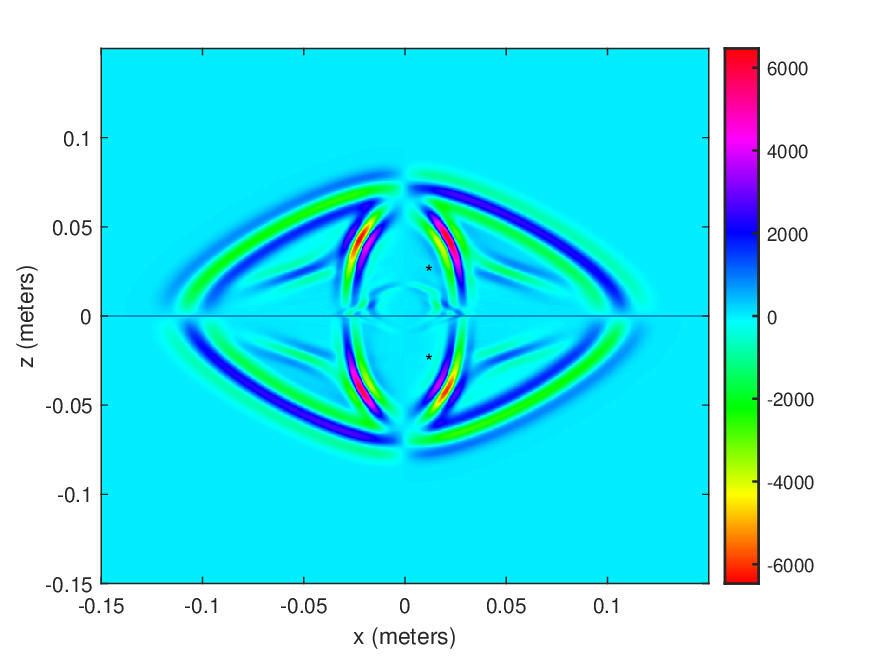}}
\subfloat[]{%
   \includegraphics[width=2.5in,height=2.0in]{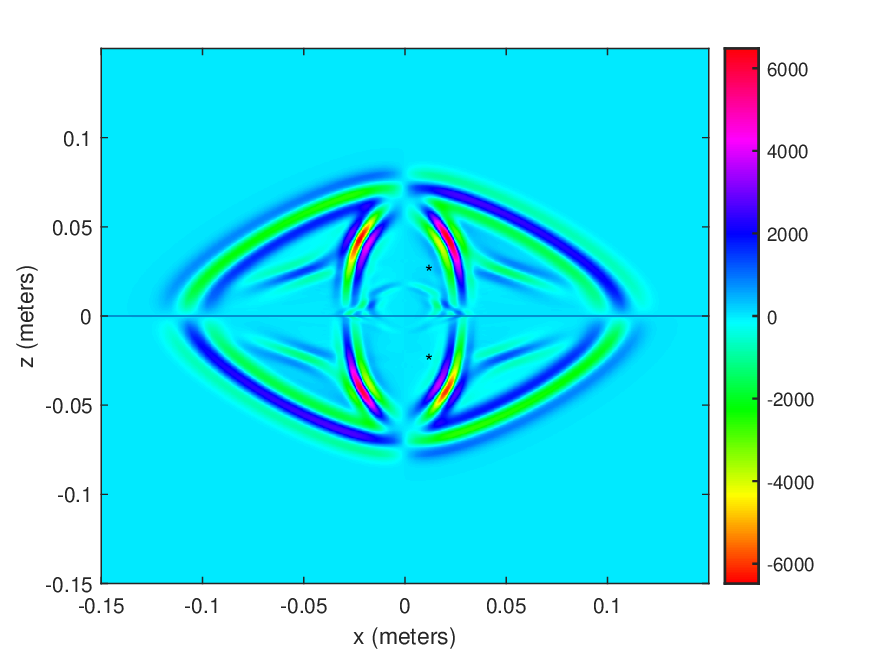}}\\
\subfloat[]{%
   \includegraphics[width=2.5in,height=2.0in]{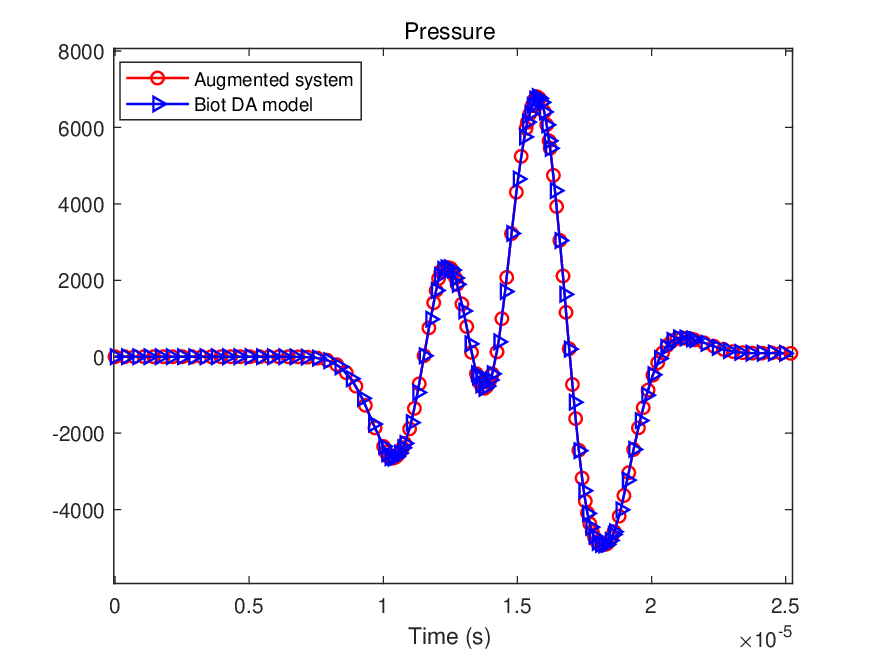}}
\subfloat[]{%
   \includegraphics[width=2.5in,height=2.0in]{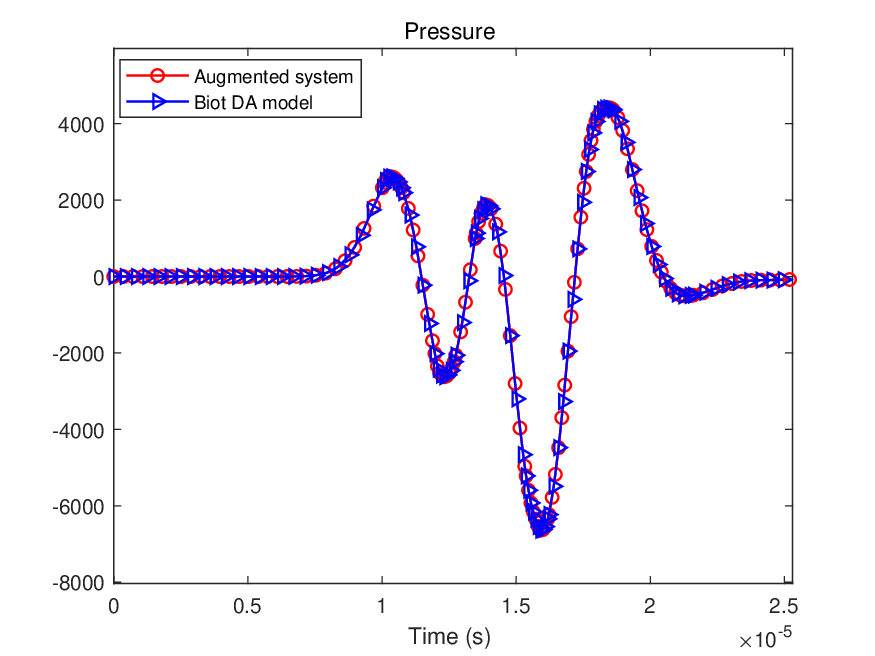}}

\caption{Pressure field of heterogeneous domain  with frequency $f_{src}$=200K Hz for  Biot-JKD equations with different methods. (a) Snapshot of pressure at $t= 2.53\times10^{-5}s$ using the Augmented Biot-JKD equations,  (b) snapshot of pressure at $t= 2.53\times10^{-5}s$ using the Biot-DA equations, (c) comparison of time evolution at the receiver $R_0$, (d) comparison of time evolution at the receiver $R_1$.}
\label{fig heterohigh}
\end{figure}

Two receivers  $R_0(x_1,z_1)=(0.01m,0.025m)$ and $R_1(x_2,z_2)=(0.01m,-0.025m)$ marked with $\star$ in Figure \ref{fig heterohigh} are selected to record the time evolution of the pressure. For comparison, we also use RKDG method to run the Biot DA model \cite{Lombard2}. Figure \ref{fig heterohigh} shows the results of the pressure $p$ with $f_{src}$= 200K Hz for the Augmented Biot-JKD system (\ref{eq:higheqn}) and the Biot-DA model, from which we can seen that the solutions from the two methods are consistent.

\section{\label{Conclude}Conclusion and future works}
In this paper, we use the RKDG method to investigate wave propagation in orthotropic poroelastic media across the full range of frequencies solving the low-frequency (LF)-Biot equations and the Biot-JKD equations, in which the frequency-dependent drag force is described by a memory term. For the LF-Biot equations, we consider both inviscid and viscous cases. For the inviscid case, accuracy is tested by computing the convergence orders between our numerical solutions and the analytic plane wave solutions, homogenous and heterogeneous examples with a point source are all considered. For the viscous case, the problem is a stiff system, and the dissipative source term is directly calculated using RKDG method with time steps that are small enough to satisfy the stability condition.  Such a way to handle the viscous term is different from the previous works \cite{Yvonne, Puente}, where the balance law is split into a propagative part and a diffusive part. For the Biot-JKD equations, the memory term is dealt with by using the technique developed in \cite{Ou-Woer} and the new system is termed the Augmented Biot-JKD equations. The poles and residues required for this method are computed from the frequency domain JKD tortuosity function with high accuracy. The simulation results we present in this paper are the very first numerical implementation for solving the Augmented Biot-JKD equations. For comparison, the fractional derivative approaches proposed in the literature are also implemented. For our test cases, the Augmented Biot-JKD equations and the fractional derivative approach give almost identical results. One of the advantages of the Augmented Biot-JKD approach is in its simplicity for implementation, another is the zero components of the flux for the auxiliary variables, which can simplify the system. Furthermore, the Augmented Biot-JKD approach has higher efficiency by avoiding discretizing the fractional derivative and storing the history solutions.

Numerical examples so far are only based on poroelastic media.  One of our future work is to expand our method to some other problems whose governing equations contain viscous effects and can be expressed as a Stieltjes function. Another is the problem of the fluid-solid interaction with a bounded transverse poroelastic medium enclosed by the unbounded homogeneous compressible inviscid fluid. In addition, we will also consider the problem that a bounded transverse poroelastic or elastic medium immerses in electromagnetic waves.

\section*{Acknowledgments}
The work of L. Xu is partially supported by a Key Project of the Major Research Plan of NSFC (No. 91630205), and a NSFC Grant (No. 11771068). J. Xie is partially supported by the project of Chongqing University 2016 graduate student research innovation project (No. CYS16017). The work of M.Y. Ou. Is partially sponsored by the US National Science Foundation grant DMS-1413039.

\end{document}